\documentclass[11pt,sort&compress]{elsarticle}
\makeatletter
\def\ps@pprintTitle{%
 \let\@oddhead\@empty
 \let\@evenhead\@empty
 \def\@oddfoot{\centerline{\thepage}}%
 \let\@evenfoot\@oddfoot}
\makeatother
\usepackage{dsfont}
\usepackage{amsmath}
\usepackage{accents}
\usepackage{amssymb}
\usepackage{mathrsfs}
\usepackage[LGR,T1]{fontenc}
\usepackage[latin1]{inputenc}
\usepackage[cal=boondox,calscaled=1]{mathalfa}
\usepackage{psfrag}
\usepackage{pstool}
\usepackage{caption}
\usepackage{graphicx}
\newcommand{\hook}{\text{\large{$\lrcorner$}}}
\usepackage{color}
\usepackage{hyperref}
\newcommand{\qq}[1]{``#1''} 
\newcommand{\utilde}[1]{\undertilde{#1}}
\newcommand{\di}{\mathrm{d}}
\usepackage{tensor}\newcommand{\ou}[3]{{#1}{}^{#2}{}_{#3}}
\newcommand{\uo}[3]{{#1}{}_{#2}{}^{#3}}

\newcommand{\uepsilon}{\boldsymbol{\underaccent{\check}{\epsilon}}}
\newcommand{\oepsilon}{\boldsymbol{\hat{\epsilon}}}
\newcommand{\hateta}{\boldsymbol{\hat{\eta}}}
\newcommand{\I}{\mathrm{i}} 
\newcommand{\E}{\mathrm{e}} 
\newcommand{\CC}{\mathrm{cc.}} 
\newcommand{\C}{\mathbb{C}}

\newcommand{\R}{\mathbb{R}}

\newcommand{\changefont}[3]{\fontfamily{#1}\fontseries{#2}\fontshape{#3}\selectfont}
\newenvironment{subalign}{\subequations\align}{\endalign\endsubequations}
\newcommand{\eref}[1]{(\ref{#1})}
\renewcommand{\b}{\bar}
\renewcommand{\bold}[1]{\boldsymbol{#1}}

\DeclareMathAlphabet{\sfit}{OT1}{fos}{sb}{it}
\DeclareMathAlphabet{\mathsf}{OT1}{fos}{sb}{n}
\newcommand{\textgreek}[1]{\begingroup\fontencoding{LGR}\selectfont#1\endgroup}
\newcommand{\sfpi}{\text{\changefont{fos}{sb}{it}\textgreek{p}}}
\newcommand{\sfPi}{\text{\changefont{fos}{sb}{n}\textgreek{P}}}

\begin{document}

\begin{abstract}
I present a model of discrete gravity, which is formulated in terms of a topological gauge theory with defects. The theory has no local degrees of freedom and the gravitational field is trivial everywhere except at a number of colliding null surfaces, which represent a system of curvature defects propagating at the speed of light. 
The underlying action is local and it is studied in both its Lagrangian and Hamiltonian formulation. The canonically conjugate variables on the null surfaces are a spinor and a spinor-valued two-surface density, which are coupled to a topological field theory for the Lorentz connection in the bulk. I discuss the relevance of the model for non-perturbative approaches to quantum gravity, such as loop quantum gravity, where similar variables have recently appeared as well. 
\end{abstract}
\title{Discrete gravity as a topological gauge theory with light-like curvature defects}
\author{Wolfgang Wieland}
\address{Perimeter Institute for Theoretical Physics\\31 Caroline Street North\\ Waterloo, ON N2L\,2Y5, Canada\\{\vspace{0.5em}\normalfont Fall 2016}
}
\maketitle{
\tableofcontents}
\begin{center}{\noindent\rule{\linewidth}{0.4pt}}\end{center}
\section{Introduction}
We will consider gravity as gauge theory for the Lorentz group. The fundamental configuration variables in the bulk are the tetrad and the self-dual connection \cite{newvariables}. For a manifold with boundaries, the action acquires boundary terms, which reconcile the variational problem with the boundary conditions. If the boundary is null, we will see that the most natural such boundary term is given by the three-dimensional integral
\begin{equation}
\int_{\partial\mathcal{M}}\bold{\pi}_A\wedge D\ell^A+\CC,
\end{equation}
where $\bold{\pi}_A$ is a spinor-valued two-form, $D\ell^A=\di\ell^A+\ou{A}{A}{B}\ell^B$ is the gauge covariant exterior derivative and $\ell^A$ 
is a spinor, whose square returns the null generators of the three-dimensional null boundary $\partial{\mathcal{M}}$.

We will then use this boundary term to discretise gravity by truncating the bulk geometries to field configurations that are locally flat. The bulk action vanishes and only the three-dimensional \emph{internal} null boundaries and two-dimensional corners contribute non-trivially. The resulting action defines a theory of distributional four-geometries in terms of a topological field theory with defects that propagate at the speed of light.  

Whether the continuum limit exists and brings us back to general relativity is a difficult question. This paper does not provide an answer. We can only give some indications in favour of our proposal: First of all, we will show that the solutions of the theory represent four-dimensional Lorentzian geometries, which are built of a network of three-dimensional null surfaces (equipped with a signature $(0$$+$$+)$ metric) glued among bounding two-surfaces. We will then find that every such null surface represents a potential curvature defect. We will also show that there exist solutions of the equations of motion derived from the discretised action that represent distributional solutions of Einstein's equations with non-vanishing Weyl curvature in the neighbourhood of a defect. In other words, the Einstein equations are satisfied locally. Whether this holds on all scales, with possible higher order curvature corrections and running coupling constants, is a more difficult question. Sophisticated coarse graining and averaging techniques, such as those developed for Regge calculus  \cite{Feinberg:1984he,Barrett:1988wd, barrett1988convergence} and related approaches cf.\ \cite{Bahr:2009mc,Bahr:2009qc} may provide useful tools for the future.

The main motivation concerns possible applications for non-\-per\-tur\-bative approaches to quantum gravity, such as loop quantum gravity \cite{ashtekar, rovelli, thiemann, status, alexreview}. In loop quantum gravity, geometry is described in terms of the inverse and densitised triad 
\begin{equation}
\uo{E}{i}{a}=\frac{1}{2}\oepsilon^{abc}\epsilon_{ilm}\ou{e}{l}{b}\ou{e}{m}{c},
\end{equation}
which is the gravity analogue of the Yang\,--\,Mills electric field. In quantum gravity, it becomes an operator, whose flux across a surface has a discrete spectrum \cite{Rovelliarea, AshtekarLewandowskiArea}. The resulting semi-classical geometry\,---\,in the naive limit, where $\hbar\rightarrow 0$, and all quantum numbers are sent to infinity, while keeping fixed the eigenvalues of geometric operators\,---\,is distributional. The semi-classical electric field 
\begin{equation}
\uo{E}{i}{a}(p)=\sum_{l}\int_l\di s\frac{\di l^a(s)}{\di s}\,E^{l}_i(s)\,\delta^{(3)}\big(p,l(s)\big)
\end{equation}
has support only along the links $l_1,\dots,l_L$ dual to a cellular decomposition of the spatial manifold. 
There is no geometry, no notion of volume, distance and area, outside of this one-dimensional fabric of space. What is then the dynamics for these distributional geometries at the quantum level? There are several proposals, such as those given by the covariant spinfoam approach \cite{reisenberger,flppdspinfoam,LQGvertexfinite,alexreview} or Thiemann's canonical program \cite{thiemann,qsd,Thiemann:2005zg}, but the precise transition amplitudes are unknown. 
We know, however, that the three-dimensional quantum geometries represent distributional excitations of the gravitational field, and we can expect, therefore, that the semi-classical $\hbar\rightarrow 0$ limit of the amplitudes will define a classical theory of discrete, or rather distributional, spacetime geometries, whose Hamiltonian dynamics is formulated in terms of gauge connection variables. 
This paper presents a proposal for such a theory, and may, therefore, open up a new road towards non-perturbative quantum gravity.


\section[Boundary spinors in GR]{Boundary spinors in general relativity}\label{sec2}

\subsection{Self-dual area two-form on a null surface}\label{sec2.1}
On a null surface $\mathcal{N}$, the pull-back of the self-dual component of the Pleba\'nski two-form $e_\alpha\wedge e_\beta$ admits a very simple algebraic description: it turns into the symmetrised tensor product \eref{fluxparam} of a spinor $\ell_A$ with a spinor-valued two-form $\bold{\eta}_A$. This is a key observation and it is important for the further development of this paper. Let us explain it in more detail.

Consider thus an oriented three-dimensional null surface $\mathcal{N}$ in a four-dimensional Lorentzian spacetime manifold\footnote{Concerning the notation, the conventions are the following: The metric signature is ($-$$+$$+$$+$) and $a,b,c,\dots$ are abstract tensor indices labelling the sections of the tensor bundle over either all of spacetime or some submanifold in there. $\alpha,\beta,\gamma, \dots$ are internal Minkowski indices. For the associated spinor bundle, we use an index notation as well: Its sections carry indices $A,B,C,\dots$ referring to the fundamental spin $(\tfrac{1}{2},0)$ representation of $SL(2,\C)$. Primed indices $\b A, \b B, \b C,\dots$ belong to the complex conjugate spin $(0,\tfrac{1}{2})$ representation. See the \hyperref[spinappdx]{appendix} for further details on the notation.} $(\mathcal{M},g_{ab})$. 
 We assume $\mathcal{M}$ to be parallizable, hence there is a spin structure and a global frame field $\ou{e}{\alpha}{a}$ diagonalising the metric $g_{ab}=\eta_{\alpha\beta}\ou{e}{\alpha}{a}\ou{e}{\beta}{b}$. We define the Pleba\'nski two-form
\begin{equation}
\Sigma_{\alpha\beta}=e_\alpha\wedge e_\beta,
\end{equation}
and split it into its self-dual and anti-self-dual components
\begin{subequations}
\begin{align}\label{selfdualforms}
\Sigma_{AB}=\frac{1}{4}\sigma_{A\b C\alpha}\ou{\b\sigma}{\b C}{B\beta}e^\alpha\wedge e^\beta,\\
\b\Sigma_{\b A\b B}=\frac{1}{4}\b\sigma_{C\b A\alpha}\ou{\sigma}{C}{\b B\beta}e^\alpha\wedge e^\beta,
\end{align}\label{selfdualcompon}%
\end{subequations}
which are built from the soldering forms $\ou{\sigma}{A\bar A}{\alpha}$ that map spinors into internal Minkowski vectors. In terms of the Dirac gamma matrices, which may be more familiar to the reader, we could also write 
\begin{equation}
\begin{pmatrix}
\ou{\Sigma}{A}{B}&\emptyset\\
\emptyset&-\uo{\b\Sigma}{\b A}{\b B}
\end{pmatrix}=-\frac{1}{8}[\gamma_\alpha,\gamma_\beta]e^\alpha\wedge e^\beta.\label{selfdualgamma}
\end{equation}
See the \hyperref[spinappdx]{appendix} for further details on the notation.

\vspace{1em}
The three-manifold $\mathcal{N}$ is null. Let then $\ell^a\in T\mathcal{N}$ denote its future pointing null generator, which is unique up to boosts $\ell^a\mapsto\E^\eta\ell^a$. We can then always find a spinor $\ell^A$, which squares to 
\begin{equation}
\ell^a=-\frac{1}{\sqrt{2}}\uo{\sigma}{A\bar A}{a}\ell^A\b\ell^{\b A}.
\end{equation}
Clearly, there is an additional gauge symmetry: given the null vector $\ell^a$, the spinor $\ell^A$ can be determined only up to residual $U(1)$ phase transformations $\ell^A\mapsto\E^{\I\phi/2}\ell^A$.

Let now $\varphi:\mathcal{N}\hookrightarrow\mathcal{M}$ be the canonical embedding of the null surface $\mathcal{N}$ into $\mathcal{M}$, and consider the pull-back $\varphi^\ast\Sigma_{AB}$ of the self-dual two-form $\Sigma_{AB}$ to $\mathcal{N}$. It always exists then a spinor-valued two-form $\bold{\eta}^A\in\Omega^2(\mathcal{N}:{\C^2})$ on $\mathcal{N}$ such that
\begin{equation}
\varphi^\ast\Sigma_{AB}=\bold{\eta}_{(A}\ell_{B)}.\label{fluxparam}
\end{equation}
This can be seen as follows: First of all we choose a second linearly independent spinor $k^A$ such that we have a normalised spin dyad $\{\ell^A,k^A\}:\epsilon_{AB}k^A\ell^B=k_A\ell^A=1$. This dyad induces a triad $\{\ell_A\ell_B,k_Ak_B,k_{(A}\ell_{B)}\}$ in the space of symmetric bispinors $S_{AB}=S_{(AB)}$, which we then use to decompose the pull-back $\varphi^\ast\Sigma_{AB}$ of the self-dual area two-form into component functions $\boldsymbol{\mu}$, $\boldsymbol{\nu}$ and $\boldsymbol{\varepsilon}$, which are complex-valued two-forms on $\mathcal{N}$. We then have
\begin{equation}
\varphi^\ast\Sigma_{AB}=\boldsymbol{\mu}\ell_A\ell_B+\boldsymbol{\nu}k_Ak_B+\I\boldsymbol{\varepsilon}k_{(A}\ell_{B)}.\label{fluxcomp}
\end{equation}
Next, we also know that $\ell^a$ is the null generator of $\mathcal{N}$. The pull-back of $\ell_a$ to $\mathcal{N}$ thus vanishes, which implies, in turn
\begin{equation}
\varphi^\ast(e_\alpha\wedge \ell)=\left(\varphi^\ast\Sigma_{\alpha\beta}\right)\ou{e}{\beta}{b}\ell^b=0.
\end{equation}
This is the same as to say
\begin{equation}
\big(\b\epsilon_{\b A\b B}\,\varphi^\ast\Sigma_{AB}+\CC\big)\ell^B\b{\ell}^{\b B}=0.
\end{equation}
We contract the free indices with all possible combinations of the spinors $\ell^A$, $k^A$ and their complex conjugate. This immediately implies that
$\boldsymbol{\nu}$ vanishes, but it also implies
the \emph{reality condition}
\begin{equation}
\boldsymbol{\varepsilon}=\b{\boldsymbol{\varepsilon}}.\label{realcond0}
\end{equation}
Since $\boldsymbol{\nu}=0$, we can go back to equation \eref{fluxcomp} and identify $\bold{\eta}_A$  with
\begin{equation}
\bold{\eta}_A=\boldsymbol{\mu}\ell_A+\I\boldsymbol{\varepsilon}k_A,\label{kappacomp}
\end{equation}
which proves the desired equation \eref{fluxparam}.
\vspace{1em}

The Lorentz invariant contraction $\bold{\eta}_A\ell^A$ has an immediate geometrical interpretation: It defines the two-form $\boldsymbol{\varepsilon}\in\Omega^2(\mathcal{N}:\R)$, which measures the oriented area of any two-dimensional spatial submanifold $\mathcal{C}$ of $\mathcal{N}$. Indeed
\begin{equation}
\pm\operatorname{Area}[\mathcal{C}]=\int_{\mathcal{C}}\boldsymbol{\varepsilon}=-\I\int_{\mathcal{C}}\bold{\eta}_A\ell^A.\label{arint}
\end{equation}
Whether this integral coincides with the metrical surface area
\begin{equation}
\operatorname{Area}[\mathcal{C}]=\int_{\mathcal{C}}\di x\,\di y\sqrt{g(\partial_x,\partial_x)g(\partial_y,\partial_y)-g(\partial_x,\partial_y)^2}
\end{equation}
depends on whether $\ell^a$, which is future pointing,  is an outgoing or incoming null generator with respect to the induced orientation\footnote{That $\mathcal{C}$ inherits an orientation from $\mathcal{N}$ is immediate. We can say, in fact, that a pair of tangent vectors $(\partial^a_x,\partial^a_y)$ in $T\mathcal{C}$ is positively oriented in $\mathcal{C}$, if the triple $(\ell^a,\partial^a_x,\partial^a_y)$ is positively oriented in $\mathcal{N}$. We can then choose an arbitrary future oriented time-like vector $t^a$, which is based on $\mathcal{C}$, and say that $\ell^a$ is outgoing with respect to $\mathcal{C}$, if the quadruple $(t^a,\ell^a,\partial^a_x,\partial^a_y)$ is positively oriented in $\mathcal{M}$, otherwise inwardly oriented, and this definition will not depend on the choice of $t^a$. A straightforward calculation shows then that the sign of $\boldsymbol{\varepsilon}(\partial_x,\partial_y)$ coincides with the orientation of $(t^a,\ell^a,\partial^a_x,\partial^a_y)$, and if $\boldsymbol{\varepsilon}(\partial_x,\partial_y)>0$, we can thus say that $\ell^a$ is an outgoing null generator with respect to $\mathcal{C}$, and it is incoming if $\boldsymbol{\varepsilon}(\partial_x,\partial_y)<0$.} on $\mathcal{C}$.  

\subsection{Boundary term on a null surface}\label{sec2.2}
The last section gave a parametrisation \eref{fluxparam} of the self-dual area two-form on a null surface $\mathcal{N}$ in terms of boundary spinors $\bold{\eta}_A$ and $\ell^A$. We now use them to build a gravitational boundary action with spinors as the fundamental boundary variables.

Working in a first-order formalism, we write the gravitational action as a functional of the Lorentz connection $\ou{A}{\alpha}{\beta}$ and the tetrad $e^\alpha$.
At the level of the action, the connection is then independent of the triad, which means that the torsionless condition does not hold \emph{off-shell}. This allows us to add the term $e_\alpha\wedge e_\beta\wedge F^{\alpha\beta}$ to the action without changing the equations of motion, which are the Einstein equations plus the torsionless condition. With the inclusion of a cosmological constant $\Lambda$, we thus work with the bulk action
\begin{align}\nonumber
S_{\mathcal{M}}[A,e]=\frac{1}{16\pi G}\int_{\mathcal{M}}&\left[\left(\ast\!\left(e_\alpha\wedge e_\beta\right)-\frac{1}{\beta}e_\alpha\wedge e_\beta\right)\wedge F^{\alpha\beta}[A]+\right.\\
&\left.\qquad\qquad-\frac{\Lambda}{6}\ast\!\left(e_\alpha\wedge e_\beta\right)\wedge e^\alpha\wedge e^\beta\right],\label{tetradactn}
\end{align}
where $\ou{F}{\alpha}{\beta}=\di\ou{A}{\alpha}{\beta}+\ou{A}{\alpha}{\mu}\wedge\ou{A}{\mu}{\nu}$ is the field strength of the Lorentz connection, and the constant in front of $e_\alpha\wedge e_\beta\wedge F^{\alpha\beta}$ is called the Barbero\,--\,Immirzi parameter $\beta$.

We then write the bulk action in terms of self-dual variables, which are the self-dual area two-form \eref{selfdualforms} and the $SL(2,\C)$ spin connection $\ou{A}{A}{B}$, whose curvature is the self-dual part of $F^{\alpha\beta}$. We have
\begin{equation}
S_{\mathcal{M}}[A,e]=\frac{\I}{8\pi \beta G}\int_{\mathcal{M}}\left[(\beta+\I)\,\Sigma_{AB}\wedge F^{AB}-\frac{\beta\Lambda}{6}\,\Sigma_{AB}\wedge\Sigma^{AB}\right]+\CC,\label{selfdualactn}
\end{equation}
where $\CC$ denotes the complex conjugate of all preceding terms. 

Consider then the variation of the action. The Einstein equations follow from the variation of the tetrad, the variation of the connection yields the torsionless condition $D\Sigma_{AB}=\di\Sigma_{AB}-2\uo{A}{(A}{C}\wedge\Sigma_{B)C}=0$ and the remainder
\begin{equation}
\frac{\I}{8\pi \beta G}(\beta+\I)\int_{\partial \mathcal{M}}\Sigma_{AB}\wedge\delta A^{AB}+\CC\label{remder}
\end{equation}
at the boundary. This boundary integral must cancel against the variation of the boundary action, otherwise the entire action is not functionally differentiable. We have assumed that the boundary $\partial \mathcal{M}$ is null, which implies that the self-dual area two-form $\Sigma_{AB}$ has the simple form of equation \eref{fluxparam}. What is then the right boundary action? The remainder \eref{remder} is linear in the connection, and we expect, therefore, that the boundary action is linear in the connection as well. Only the exterior covariant derivative $D=\di+[A,\cdot]$ is both gauge covariant and linear in $\ou{A}{A}{Ba}$, which suggests that the boundary term is built from the gauge covariant derivative of some boundary fields. The only available fields at the boundary, which are functionally independent of the connection, are the boundary spinors $\bold{\eta}_A$ and $\ell^A$ themselves. 
Now, $\bold{\eta}_A$ is a two-form, such that $\bold{\eta}_A\wedge D\ell^A$ and $\bar{\bold{\eta}}_{\bar A}\wedge D\bar\ell^{\bar A}$ are complex-valued three-forms, whose boundary integrals define the most obvious candidates for the gravitational boundary term at a null surface (expressions like $D\bold{\eta}_A\wedge\ell^A$ are related to the latter by a total derivative). We are left to determine the coupling constants in front, which can be read off the remainder \eref{remder} of the connection variation at the boundary. The resulting boundary term is
\begin{equation}
\frac{\I}{8\pi\beta G}(\beta+\I)\int_{\partial\mathcal{M}}\bold{\eta}_A\wedge D\ell^A+\CC,\label{boundactn}
\end{equation}
where $D_a$ denotes the gauge covariant derivative
\begin{equation}
D_a\ell^A=\partial_a\ell^A+\ou{A}{A}{Ba}\ell^B.\label{covderiv}
\end{equation}
 
Notice that there is now an additional $U(1)_{\C}$ gauge symmetry appearing: The spinors $\bold{\eta}_A$ and $\ell^A$ are unique only up to transformations
\begin{equation}
(\bold{\eta}_A,\ell^A)\longrightarrow(\E^{\frac{z}{2}}\bold{\eta}_A,\E^{-\frac{z}{2}}\ell^A),
\end{equation}
where the gauge element $z:\mathcal{N}\rightarrow\C$ generates both boost and rotations preserving the null normal $\I\ell^A\bar\ell^{\bar A}$. The boundary action \eref{boundactn} is not invariant under this symmetry, but we can easily make it invariant by introducing a fiducial $U(1)_{\C}$ gauge connection $\omega$ and writing
\begin{equation}
S_{\partial\mathcal{M}}[A|\bold{\pi},\ell,\omega]=\int_{\partial\mathcal{M}}\bold{\pi}_A\wedge (D-\omega)\ell^A+\CC,\label{bndryactn0}
\end{equation}
where we introduced the momentum spinor $\bold{\pi}_A$ as an abbreviation for
\begin{equation}
\bold{\pi}_A:=\frac{\I}{8\pi\beta G}(\beta+\I)\bold{\eta}_A.\label{pispindef}
\end{equation}
The introduction of an additional $U(1)_\C$ connection $\omega$ to render \eref{bndryactn0} fully gauge invariant may seem rather dull, but it will be important for us later, when we will learn how to glue causal regions across a bounding null surface. 
 
Notice also, that the spinors $\bold{\pi}_A$ and $\ell^A$ are not independent, for we have to satisfy the reality conditions \eref{realcond0}, which now turn into
\begin{equation}
\frac{\I}{\beta+\I}\bold{\pi}_A\ell^A+\CC=0.\label{realcond}
\end{equation}
If \eref{realcond} is satisfied, the area is real and we can define the oriented area two-form on a null surface $\mathcal{N}$ simply by
\begin{equation}
\boldsymbol{\varepsilon}=-\frac{8\pi\beta G}{\beta+\I}\bold{\pi}_A\ell^A.\label{areaform}
\end{equation}
The entire action for the gravitational degrees of freedom in a region $\mathcal{M}$ bounded by a null surface $\mathcal{N}$ is therefore given by the expression
\begin{equation}
S[A,e|\bold{\pi},\ell,\omega]=S_{\mathcal{M}}[A,e]+S_{\mathcal{N}}[A|\bold{\pi},\ell,\omega].
\end{equation}
At this point, the boundary spinors $\bold{\pi}_A$ and $\ell^A$ are not varied in the action, they are kept fixed in the variational principle, because they determine the boundary value \eref{fluxparam} of the self-dual two-form $\varphi^\ast\Sigma_{AB}$ as an equation of motion derived from the variation of the connection in the bulk and boundary.

In the literature, other boundary terms have been used on null surfaces as well. For a recent survey in the metric formalism, we refer to \cite{LehnerBoundary} and references in there. The boundary term \eref{bndryactn0} is a generalisation\,---\,it is formulated in terms of spinors, and does not assume that the connection is torsionless, which explains the implicit appearance of the Barbero\,\---\,Immirzi parameter, which enters the action \eref{bndryactn0} through the definition of the momentum spinor \eref{pispindef}.  Notice also that additional corner terms may be necessary as well. We will introduce them below.


\section{Discretised gravity with impulsive gravitational waves}\label{sec3}
\subsection{Glueing flat four-volumes along null surfaces}\label{sec3.1}
In Regge calculus \cite{Reggecalc}, the Einstein equations are discretised by cutting the spacetime manifold $\mathcal{M}$ into four-simplices, and truncating the metric to field configurations that are locally flat. The gravitational action for the entire manifold is then a sum over all such four-simplices, each one of which contributes a bulk and boundary term. 
 The bulk contribution vanishes (for $\Lambda=0$), and we are left with a distributional boundary term, which can be reorganised into a sum over triangles, with every triangle contributing its area times the surrounding deficit angle.

The task is then to generalise Regge calculus and find a theory of discretised gravity in the connection formalism, whose action is still simple enough to admit a Hamiltonian quantisation. We will propose such a theory by dropping the assumption that the elementary building blocks are flat four-simplices. We work instead with four-dimensional regions, which are flat or constantly curved inside (depending on the value of the cosmological constant), and whose boundary is null. 

The theory is then specified by the matching conditions that determine the discontinuity of the gravitational field in the vicinity of the interjacent null surface. In Regge calculus, it is the intrinsic three-dimensional geometry at the interface that is matched between the two sides. We require this condition as well, and thus impose that
\begin{equation}
q_{ab}=\utilde{q}_{ab},\label{metrmatch}%
\end{equation}%
where $q_{ab}$ denotes the intrinsic three-metric from below\footnote{The null surface $\mathcal{N}$ is oriented, and the null vectors $\ell^a$ are future pointing. In a neighbourhood of $\mathcal{N}$ we can thus distinguish points sitting below the null surface from those lying above. The quantities $q_{ab}, \ell^a, \ell^A, \bold{\eta}_A, k_A,\dots$ describe the boundary as seen from below, the tilde quantities $\utilde{q}_{ab}, \utilde{\ell}^a, \utilde{\ell}^A, \utilde{\bold{\eta}}_A,\utilde{k}_A,\dots$ describe the boundary from above.} the interjacent null surface $\mathcal{N}$, while $\utilde{q}_{ab}$ determines the geometry from the other side. 

In our case, the interface is null, and the induced three-metric $q_{ab}=\varphi^\ast g_{ab}$ is degenerate. It has signature $(0$$+$$+)$, and the null vector $\ell^a$ defines the single degenerate direction $\ell^a:q_{ab}\ell^b=0$. 
There is then also $\utilde{q}_{ab}$ with null vector $\utilde{\ell}^a$, and equation \eref{metrmatch} implies that they match from the two sides, i.e.\ $\ell^a\sim\utilde{\ell}^a$, which is the same as to say
\begin{equation}
\exists\eta:\mathcal{N}\rightarrow\R:\ell^a=\E^\eta\utilde{\ell}^a.\label{nullgenmatch}
\end{equation}

In our formalism, the fundamental configuration variables are the boundary spinors $\ell^A$ and $\bold{\eta}_A$ rather than the metric $q_{ab}$ and the null vector $\ell^a$. How do we then express the glueing conditions \eref{matchcond} in terms of the new variables? The conditions $\ell^A=\utilde{\ell}^A$, $\bold{\eta}_A=\utilde{\bold{\eta}}_A$ would certainly be sufficient, but they are too strong, for they also match unphysical gauge degrees of freedom, which are absent in \eref{metrmatch}. We should thus only match $SL(2,\C)$ gauge invariant combinations of $\ell^A$ and $\bold{\eta}_A$. In other words, we ought to impose
\begin{subequations}
\begin{align}
\uo{\boldsymbol{\eta}}{A}{a}{\ell}^A&=
\uo{\utilde{\boldsymbol{\eta}}}{A}{a}\utilde{\ell}^A\label{areamatch},\\
\uo{\boldsymbol{\eta}}{A}{a}\boldsymbol{\eta}^{Ab}&=\uo{\utilde{\boldsymbol{\eta}}}{A}{a}\utilde{\boldsymbol{\eta}}^{Ab},\label{shapematch}
\end{align}\label{matchcond}%
\end{subequations}
where the spinor-valued vector-density $\uo{\boldsymbol{\eta}}{A}{a}$ denotes the densitised dual of the two-form $\ou{\bold{\eta}}{A}{ab}$, namely
\begin{equation}
\uo{\boldsymbol{\eta}}{A}{a}:=\frac{1}{2}\oepsilon^{abc}\bold{\eta}_{Abc},
\end{equation}
and $\oepsilon^{abc}$ is the metric-independent Levi-Civita density on $\mathcal{N}$, which is defined for any three-dimensional coordinate system $\{x^i\}$ on $\mathcal{N}$ simply by $\oepsilon^{abc}=\di x^i\wedge\di x^j\wedge \di x^k\partial_i^a\partial_j^b\partial_k^c$.

\paragraph{Area-matching condition} We now have to convince ourselves that the matching conditions for the spinors \eref{matchcond} are indeed equivalent to the conditions \eref{metrmatch} and \eref{nullgenmatch} for $\ell^a$ and $q_{ab}$. We start with the \emph{area-matching condition} \eref{areamatch}. Going back to equation \eref{kappacomp} we decompose both $\bold{\eta}_A$ and $\utilde{\bold{\eta}}_A$ into a normalised\footnote{We can always extend $\ell^A$ (resp.\ $\utilde{\ell}^A$) with a second linearly independent spinor $k^A$ (resp.\ $\utilde{k}^A$) into a local spin basis such that $k_A\ell^A=1=\utilde{k}_A\utilde{\ell}^A$.} spin dyad $\{k^A,\ell^A\}$ and $\{\utilde{k}^A,\utilde{\ell}^A\}$ on either side, and write
\begin{subequations}\begin{align}
\bold{\eta}_A&=\bold{\mu}\ell_A+\I\bold{\varepsilon} k_A,\label{bothcomps1}\\
\utilde{\bold{\eta}}_A&=\utilde{\bold{\mu}}\utilde{\ell}_A+\I\utilde{\bold{\varepsilon}}\utilde{k}_A\label{bothcomps2},
\end{align}\label{bothcomps}\end{subequations}%
where $\bold{\mu}$ and $\bold{\varepsilon}$ and $\utilde{\bold{\mu}}$ and $\utilde{\bold{\varepsilon}}$ are the component functions of $\bold{\eta}_A$ and $\utilde{\bold{\eta}}_{A}$ as in \eref{kappacomp} above. We contract both \eref{bothcomps1} and \eref{bothcomps2} with $\ell^A$ and $\utilde{\ell}^A$, going back to \eref{areamatch} we then obtain the \emph{area-matching condition}
\begin{equation}
\boldsymbol{\varepsilon}_{ab}=\utilde{\boldsymbol{\varepsilon}}_{ab}.\label{armatchvar}
\end{equation}
We then also know from the reality condition \eref{realcond0} that the area two-form $\boldsymbol{\varepsilon}\in\Omega^2(\mathcal{N}:\C)$ must be real and so is $\utilde{\boldsymbol{\varepsilon}}$. Since $\boldsymbol{\varepsilon}_{ab}$ is a real-valued two-form in three-dimensions, it has at least one degenerate direction. We assume it has only one, because otherwise $\boldsymbol{\varepsilon}_{ab}$ would vanish identically. This single degenerate eigenvector of $\boldsymbol{\varepsilon}_{ab}$ (resp. $\utilde{\boldsymbol{\varepsilon}}_{ab}$) defines the direction of the null generators $\ell^a$ (resp. $\utilde{\ell}^a$), and the matching condition \eref{areamatch} implies that they both point into the same direction, hence $\ell^a\sim\utilde{\ell}^a$ as desired.

Having shown that the area-matching condition \eref{areamatch} implies the matching \eref{nullgenmatch} of the null vectors, we are now left to show that the \emph{shape-matching conditions} \eref{shapematch} are equivalent to the remaining glueing conditions \eref{metrmatch} for the intrinsic three-metric $q_{ab}=\varphi^\ast g_{ab}$ on the two sides.
This requires some preparation: First of all, we have to understand how to reconstruct the three-metric $q_{ab}=\varphi^\ast g_{ab}$ from the spinors $\bold{\eta}_{Aab}$ and 
$\ell^A$ alone.

\paragraph{Reconstruction of $q_{ab}$ from $\boldsymbol{\eta}_A$ and $\ell^A$} To reconstruct the three-metric from the spinors, it is more intuitive to work with densitised vectors rather than three-forms on $\mathcal{N}$. Hence, we fix a fiducial volume element $\hateta\in\Omega^3(\mathcal{N}:\R)$ on $\mathcal{N}$, which is the same from the two sides $\hateta=\utilde{\hateta}$ and dualise the component functions \eref{bothcomps} $\boldsymbol{\mu}$ and $\boldsymbol{\varepsilon}$ of the spinor-valued two-form $\bold{\eta}_A=\bold{\mu}\ell_A+\I\bold{\varepsilon}k_A$. We can then always rescale the null generator $\ell^a$ such that
\begin{equation}
\frac{1}{2}\oepsilon^{abc}\boldsymbol{\varepsilon}_{bc}=\hateta\ell^a,\label{elldensity}
\end{equation}
and equally for $\utilde{\ell}^a$. We assume $\ell^a\neq 0$, otherwise the geometry would be degenerate. We then also have the component two-form $\boldsymbol{\mu}\in\Omega^2(\mathcal{N}:\C)$. Its densitised dual defines a tangent vector $\mu^a$ in the complexified tangent space $(T\mathcal{N})_\C$ through
\begin{equation}
\frac{1}{2}\oepsilon^{abc}\boldsymbol{\mu}_{bc}=:-\I\boldsymbol{\hat{\eta}}\bar{\mu}^a,\label{mudensity}
\end{equation}
equally for $\utilde{\mu}^a\in T\mathcal{N}$ from the other side of $\mathcal{N}$. What is then the geometric interpretation of $\mu^a$? The answer is simple: It defines a dual dyad $\{m_a,\bar{m}_a\}$, which diagonalises the intrinsic three-metric $q_{ab}=\varphi^\ast g_{ab}$ as $q_{ab}=2m_{(a}\bar{m}_{b)}$. This can be seen as follows: We define the complex-valued one-form $m_a$ on $\mathcal{N}$ by lowering an index with $\bold{\varepsilon}_{ab}$. Hence we say
\begin{equation}
m_a:=\I N\mu^b\boldsymbol{\varepsilon}_{ba}.
\end{equation}
The normalisation $N$ will be determined in a moment. The co-vector $m_a$ may vanish, but this is a singular case. It implies $\mu^a\propto\ell^a$, which is the same as to say that there exists a normalised spin dyad $\{k^A,\ell^A\}$ such that the pull-back of the self-dual two-form to $\mathcal{N}$, i.e.\ $\varphi^\ast \Sigma_{AB}$, assumes the form $\varphi^\ast \Sigma_{AB}=\I\boldsymbol{\varepsilon}k_{(A}\ell_{B)}$. This defines a degenerate geometry, because  the null-surface $\mathcal{N}$ becomes effectively two-dimensional\,---\,it has no affine extension along its null generators $\ell^a$. The case where $\{m_a,\bar{m}_a\}$ is linearly dependent (but $m_a\neq 0$) is equally degenerate as well, for it implies that the triple $\{\ell^a,\mu^a,\bar{\mu}^a\}$ is linearly dependent. It then follows from $\bold{\mu}_{ab}\bar\mu^b=0$ that $\boldsymbol{\mu}_{ab}\mu^a\ell^b=0$ and hence also $(\varphi^\ast\Sigma^{\alpha\beta})_{ab}\mu^a\ell^b=0$, where $\varphi^\ast$ denotes the pull-back to $\mathcal{N}$. This is incompatible with the existence of a non-degenerate tetrad in the neighbourhood of $\mathcal{N}$, for $(\varphi^\ast\Sigma^{\alpha\beta})_{ab}\mu^a\ell^b=0$ would imply $(\varphi^\ast e^{[\alpha})_a(\varphi^\ast e^{\beta]})_b\mu^a\ell^b=0$, hence $(\varphi^\ast e^\alpha)_a\ell^a\propto(\varphi^\ast e^\alpha)_a\mu^a$, which is incompatible with $\ou{e}{\alpha}{a}$ being invertible around $\mathcal{N}$.

The generic case, where the triple $\{\ell^a,\mu^a,\bar{\mu}^a\}$ is linearly independent in $(T^\ast\mathcal{N})_\C$, corresponds to a non-degenerate tetrad $\varphi^\ast e^\alpha=\I/\sqrt{2}\uo{\sigma}{\alpha}{A\bar A}\varphi^\ast e^{A\bar A}$ on $\mathcal{N}$. To reconstruct this tetrad, we proceed as follows: First of all, we fix the normalisation $N$ of $m_a$ by demanding
\begin{equation}
\bar{m}_a{\mu}^a=\pm1.\label{sgn1}
\end{equation}
The sign depends on the orientation of $\{\ell^a,\mu^a,\bar{\mu}^a\}$ with respect to the fiducial volume form, defined as
\begin{equation}
\mathrm{sgn}\big(\I\hateta_{abc}\ell^a\mu^b\bar{\mu}^c\big)=\pm 1.\label{sgn2}
\end{equation}
Next, we extend $\{m_a,\bar{m}_a\}$ with a third linearly independent co-vector $k_a=\bar{k}_a\in T^\ast\mathcal{N}$ into a dual basis of $(T^\ast\mathcal{N})_\C$, such that
\begin{equation}
\ell^a k_a=\mp1,\quad \mu^a k_a=0,\label{sgn3}
\end{equation}
with all signs in \eref{sgn1}, \eref{sgn2} and \eref{sgn3} matching according to the indicated pattern. We then have the additional freedom to perform the rescaling $(\ell^a,k_a)\rightarrow(\E^\eta\ell^a,$ $\E^{-\eta}k_a)$ for a boost angle $\eta$. We remove this ambiguity by demanding that the fiducial volume element equals
\begin{equation}
\hateta=\pm\I\oepsilon^{abc}k_a\bar{m}_bm_c.
\end{equation}
This allows us to write the pull-back of the tetrad to $\mathcal{N}$ as\begin{equation}
(\varphi^\ast e^{A\bar A})_a=\mp\I\ell^A\bar{\ell}^{\bar A}k_a\pm\I\ell^A\bar{k}^{\bar A}\bar{m}_a\pm\I k^A\bar{\ell}^{\bar A}m_a.\label{pulltetra}
\end{equation}
A short calculation reveals that this parametrisation is indeed compatible with the boundary spinors $\bold{\eta}_A=\boldsymbol{\mu}\ell_A+\I\boldsymbol{\varepsilon} k_A$ and $\ell^A$ on the null surface: Given \eref{pulltetra}, we compute the pull-back of the self-dual two-form to $\mathcal{N}$, and get
\begin{align}\nonumber
(\varphi^\ast\Sigma_{AB})_{ab}&=(\varphi^\ast\uo{e}{A}{\bar C})_{[a}(\varphi^\ast e_{B\bar C})_{ b]}=\\
&=+2\ell_A\ell_Bk_{[a}\bar{m}_{b]}+2\ell_{(A}k_{B)}\bar{m}_{[a}m_{b]},
\end{align}
which agrees with the component functions $\boldsymbol{\mu}_{ab}$ and $\boldsymbol{\varepsilon}_{ab}$ of $\bold{\eta}_{Aab}$, as written in \eref{elldensity} and \eref{mudensity}, since indeed
\begin{subequations}\begin{align}
\hateta\bar{\mu}^a&=\I\oepsilon^{abc}k_{b}\bar{m}_{c},\\
\hateta\ell^a&=\I\oepsilon^{abc}m_b\bar{m}_c.\label{eparam}
\end{align}\end{subequations}
By duality, equation \eref{eparam} is the same as to say that the area element is the wedge product 
\begin{equation}
\boldsymbol{\varepsilon}_{ab}=2\I m_{[a}\bar{m}_{b]},\label{volelmnt}
\end{equation}
of the two-dimensional co-dyad $\{m_a,\bar{m}_a\}$. 
We can then, finally, also compute the induced metric $q_{ab}=(\varphi^\ast e_{A\bar A})_a(\varphi^\ast e^{A\bar A})_b$. A short calculation gives
\begin{equation}
q_{ab}=2m_{(a}\bar{m}_{b)},\label{threemetrc}
\end{equation}
which concludes the reconstruction of the induced geometry of the null surface $\mathcal{N}$ from the boundary spinors $\bold{\eta}_A$ and $\ell^A$ alone.

\paragraph{Shape-matching conditions}  We are now left to show that the shape-matching conditions \eref{shapematch} imply that the intrinsic three-metrics $q_{ab}$ and $\utilde{q}_{ab}$ match between the two sides. To show this, we first extend $\ell^A$ (resp.\ $\utilde{\ell}^A$) with a second linearly independent spinor $k^A$ (resp. $\utilde{k}^A$) into a normalised basis $\{k^A,\ell^A\}:k_A\ell^A=1$ (resp.\ $\utilde{k}_A\utilde{\ell}^A=1$) on the two sides.
We then have the decomposition \eref{bothcomps} of the spinor-valued two-form $\bold{\eta}_A=\bold{\mu}\ell_A+\I\bold{\varepsilon}k_A$ (resp.\ $\utilde{
\bold{\eta}}_A$) into the normalised basis spinors, which brings the glueing conditions \eref{shapematch} into the form
\begin{align}\nonumber
\uo{\boldsymbol{\eta}}{A}{a}\boldsymbol{\eta}^{Ab}&=\hateta^2\big(-\I\ell_A\bar{\mu}^a+\I k_A\ell^a\big)\big(-\I\ell^A\bar{\mu}^b+\I k^A\ell^b\big)=\\
&=2\hateta^2\ell^{[a}\bar{\mu}^{b]}\stackrel{!}{=}2\hateta^2\utilde{\ell}^{[a}\utilde{\bar\mu}^{b]},\label{shapematchvar}
\end{align}
where we have used the same fiducial three-volume $\hateta$ as in \eref{elldensity} above. Having introduced a fixed fiducial volume element, we can remove the density weights and the area-matching condition \label{armatchvar} boils down to $\ell^a=\utilde{\ell}^a$, which is a consequence of \eref{elldensity}. Going back to \eref{shapematchvar}, we thus have
 \begin{equation}
\uo{\boldsymbol{\eta}}{A}{a}\boldsymbol{\eta}^{Ab}=\uo{\utilde{\boldsymbol{\eta}}}{A}{a}\utilde{\boldsymbol{\eta}}^{Ab}\Leftrightarrow\exists \zeta:\mathcal{N}\rightarrow\C:\utilde{\mu}^a=\mu^a+\zeta\ell^a.
\end{equation}
This shows that the spinor valued two-from $\utilde{\bold{\eta}}^A\in\Omega^2(\mathcal{N}:\C^2)$ admits the decomposition
\begin{equation}
\utilde{\bold{\eta}}_{Aab}=(\boldsymbol{\mu}_{ab}-\I\boldsymbol{\varepsilon}_{ab}\bar{\zeta})\utilde{\ell}_A+\I\boldsymbol{\varepsilon}_{ab}\utilde{k}_A.
\end{equation}
We can now replace $\utilde{k}_A$ by
\begin{equation}
\utilde{k}_A-\bar{\zeta}\utilde{\ell}_A,
\end{equation}
without actually changing the canonical normalisation $\utilde{k}_A\utilde{\ell}^A=1$ of the spin dyad $\{\utilde{k}^A,\utilde{\ell}^A\}$. If the glueing conditions \eref{matchcond} are satisfied, we have thus shown that there always exists normalised spin dyads $\{k^A,\ell^A\}$, $\{\utilde{k}^A,\utilde{\ell}^A\}$ on either side of the interface, such that
\begin{subalign}
\bold{\eta}_{Aab}&=\boldsymbol{\mu}_{ab}\,\ell_A+\I\boldsymbol{\varepsilon}_{ab}\,k_A,\\
\utilde{\bold{\eta}}_{Aab}&=\boldsymbol{\mu}_{ab}\,\utilde{\ell}_A+\I\boldsymbol{\varepsilon}_{ab}\,\utilde{k}_A.
\end{subalign}
In other words, there are always spin dyads $\{k^A,\ell^A\}$ and $\{\utilde{k}^A,\utilde{\ell}^A\}$ such that the component functions $\boldsymbol{\mu}$, $\boldsymbol{\varepsilon}$ and $\utilde{\boldsymbol{\mu}}$, $\utilde{\boldsymbol{\varepsilon}}$ of the spinor valued two-forms $\bold{\eta}_A$ and $\utilde{\bold{\eta}}_A$ are the same on the two sides. But now we also know that the intrinsic three-geometry $q_{ab}$ is already uniquely determined by the component functions $\boldsymbol{\mu}$, $\boldsymbol{\varepsilon}$ of the boundary spinors (as shown in the derivation of \eref{threemetrc} above). If the component functions $\boldsymbol{\mu}$, $\boldsymbol{\varepsilon}$ and $\utilde{\boldsymbol{\mu}}$, $\utilde{\boldsymbol{\varepsilon}}$ agree on the two sides, the resulting three-metrics $q_{ab}$ and $\utilde{q}_{ab}$ must agree as well. This concludes the argument, for it implies that the intrinsic three-metric is the same whether we compute it from the spinors on one side or the other. In other words
\begin{equation}
q_{ab}=\utilde{q}_{ab},
\end{equation}
which is the desired constraint \eref{metrmatch} as derived from both the area-matching and shape-matching constraints \eref{areamatch} and \eref{shapematch}.
Notice also that the map from $(\bold{\eta}_{A},{\ell}^A)$ to $(\utilde{\bold{\eta}}_{A},\utilde{\ell}^A)$ is a Lorentz transformation, which is given explicitly by
\begin{equation}
\ou{h}{A}{B}=\utilde{\ell}^Ak_B-\utilde{k}^A\ell_B:\mathcal{N}\rightarrow SL(2,\C),\label{linkhol}
\end{equation}
such that
\begin{subequations}
\begin{align}
\utilde{\ell}^A&=\ou{h}{A}{B}\ell^B,\\
\utilde{\bold{\eta}}^A&=\ou{h}{A}{B}\bold{\eta}^B.
\end{align}\label{transfunctns}
\end{subequations}

Before we go on to the next section, let me briefly summarise: In this section, we have studied the conditions to glue two adjacent regions along a null surface $\mathcal{N}$. In terms of metric variables, equations \eref{nullgenmatch} and \eref{metrmatch} match the null generator and the intrinsic three-dimensional metric $q_{ab}=\varphi^\ast g_{ab}$ from the two sides. We then saw in \eref{metrmatch} how to write these equations in terms of the boundary spinors $\bold{\eta}_A$ and $\ell^A$. There are two kinds of constraints: The area-matching condition \eref{areamatch} and the shape-matching condition \eref{shapematch}. The terminology should be clear: Equation \eref{areamatch} matches the two-dimensional area elements $\boldsymbol{\varepsilon}_{ab}$ and $\utilde{\boldsymbol{\varepsilon}}_{ab}$ from the two sides, while the shape-matching conditions \eref{shapematch} imply that all angles drawn on $\mathcal{N}$ are the same whether we compute them from the boundary spinors on one side or the other. Notice also, that the number of constraints is the same for both variables: The induced three-metric $q_{ab}$ has signature $(0$$+$$+)$, hence there are five independent matching constraints in the metric formalism. 
In terms of spinors we have five constraints as well. The reality conditions \eref{realcond} $\boldsymbol{\varepsilon}_{ab}=\bar{\boldsymbol{\varepsilon}}_{ab}$ reduce the area-matching  constraints \eref{areamatch} to three real constraints. The shape-matching conditions \eref{shapematch}, on the other hand, add only one additional complex constraint. This is not obvious from equation \eref{shapematch},  but it is immediate when we look at \eref{shapematchvar}. In both formalisms, we are thus dealing with the same five number of constraints.
\subsection{Definition of the action}\label{sec3.2}

\begin{figure}[h]
\begin{center}
\psfrag{E}{$\mathcal{M}_2$}
\psfrag{B}{\small$\mathcal{C}_{12}^{34}$}
\psfrag{D}{$\mathcal{M}_1$}
\psfrag{A}{\small$\mathcal{N}_{13}$}
\psfrag{C}{\small$\mathcal{N}_{42}$}
\psfrag{G}{$\mathcal{M}_4$}
\psfrag{F}{$\mathcal{M}_3$}
\psfragfig[scale=1]{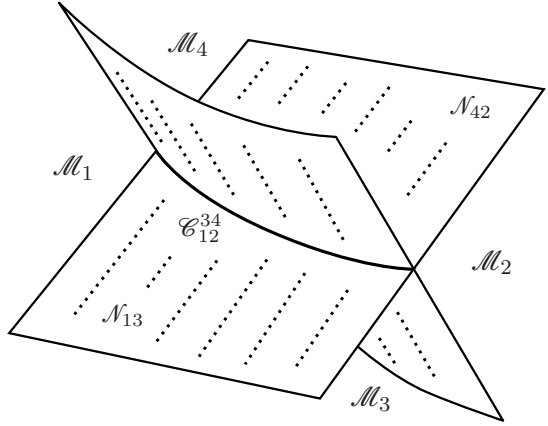}
\end{center}
\caption{The spacetime manifold $\mathcal{M}$ splits into a union of four-dimensional cells $\mathcal{M}_1,\mathcal{M}_2,\dots$, which contain no local degrees of freedom inside. Non-trivial curvature is confined to three-dimensional interfaces $\mathcal{N}_{13}=\mathcal{M}_1\cap\mathcal{M}_3,\dots$, which intersect in two-dimensional corners $\mathcal{C}_{12}^{34},\dots$.}\label{fig1}
\end{figure}

\paragraph{General idea} To now discretise gravity with the new boundary variables $\bold{\pi}_A$ and $\ell^A$ (as introduced in e.g. equation \eref{bndryactn0} above), we first introduce a cellular decomposition and cut the four-dimensional oriented manifold $\mathcal{M}$ into a finite family\footnote{The orientation of every $\mathcal{M}_i$ matches the orientation of $\mathcal{M}$, and every $\mathcal{M}_i$ is homeomorphic to a closed four-ball in $\R^4$.} of closed cells $\{\mathcal{M}_1,\mathcal{M}_2,\dots\mathcal{M}_N\}$, which are flat or constantly curved inside, i.e.
\begin{equation}
\forall p\in\mathcal{M}_i:F_{AB}(p)-\frac{\Lambda}{3}\Sigma_{AB}(p)=0.\label{Lflat}
\end{equation}
We require, in addition, that the intersection of any two such regions $\mathcal{M}_i$ and $\mathcal{M}_{j}$ is at most three-dimensional. If it is three-dimensional, we give it a name and call $\mathcal{M}_i\cap\mathcal{M}_j=:\mathcal{N}_{ij}$ an interface, whose orientation is chosen so as to match the induced orientation from $\mathcal{M}_i$. In other words $\mathcal{N}_{ij}^{-1}=\mathcal{N}_{ji}$. If, on the other hand, $\mathcal{M}_i$ and $\mathcal{M}_j$ intersect in a two-dimensional surface, we call it a corner $\mathcal{C}$, and we shall also assume, for further consistency, that all corners in the interior of $\mathcal{M}$ are adjacent to four definite such regions\,---\,four and not three or five, simply because we require that the internal boundaries are null, in which case all such corners arise from the intersection of two such null surfaces. See \hyperref[fig1]{figure 1} for an illustration.

The requirement that the boundary $\partial\mathcal{M}_i=\bigcup_{j}\mathcal{N}_{ij}$ of all four-di\-mensional building blocks is null, allows us then to use the spinors $(\bold{\eta}_A,\ell^A)$ as boundary variables: For any such internal boundary $\mathcal{N}_{ij}$ there exists a spinor\footnote{The vertical position of the $(ij)$-indices has no geometrical significance. Simplifying our notation, we thus write $\ell^A_{ij}=\epsilon^{AB}\ell^{ij}_{B}$ or, wherever possible, drop the $(ij)$-indices altogether.} $\ell^A_{ij}:\mathcal{N}_{ij}\rightarrow\C^2$ and a spinor-valued two-form $\bold{\eta}^A_{ij}\in\Omega^2(\mathcal{N}_{ij}:\C^2)$, such that the pull-back of the self-dual area two-form $\Sigma_{AB}$ (as in \eref{selfdualforms}) admits the decomposition
\begin{equation}
\varphi^{\ast}_{ij}\Sigma_{AB}^{ij}=\bold{\eta}_{(A}^{ij}\ell_{B)}^{ij},\label{boundarymatch}
\end{equation}
where $\varphi_{ij}:\mathcal{N}_{ij}\hookrightarrow\mathcal{M}$ is the canonical embedding of $\mathcal{N}_{ij}$ into $\mathcal{M}$. That this is the same as to say that $\mathcal{N}_{ij}$ is null, has been shown in section \ref{sec2.1} above following the discussion of equation \eref{fluxparam}. 

The discontinuity of the metric across the null surface will be encoded in a discontinuity of the spinors and the connection. Along a given null surface $\mathcal{N}_{ij}$, we will have two kinds of spinors, namely $(\bold{\pi}_A^{ij},\ell_{ij}^A)$ and $(\bold{\pi}_A^{ji},\ell_{ji}^A)$ from either $\mathcal{M}_{i}$ or $\mathcal{M}_{j}$. Equally for the connection: $\ou{[A^{ij}]}{A}{Ba}$ denotes the pullback of the $SL(2,\C)$ connection from the bulk $\mathcal{M}_i$ to the boundary component $\mathcal{N}_{ij}\subset\partial\mathcal{M}_i$, and $\ou{[A^{ji}]}{A}{Ba}$ is the pull-back from $\mathcal{M}_j$ to $\mathcal{N}_{ji}\subset\partial\mathcal{M}_j$. The common interface $\mathcal{N}_{ij}$ between $\mathcal{M}_i$ and $\mathcal{M}_j$ carries then two independent $SL(2,\C)$ connections $\ou{[A^{ij}]}{A}{Ba}$ and $\ou{[A^{ji}]}{A}{Ba}$. What is the relation between the two? Consider first the $SL(2,\C)$ transformation $\ou{[h^{ij}]}{A}{B}$,
\begin{subequations}
\begin{align}
{\ell}^A_{ij}&=\ou{[h^{ji}]}{A}{B}\ell^B_{ji},\\
{\bold{\pi}}^A_{ij}&=\ou{[h^{ji}]}{A}{B}\bold{\pi}^B_{ji},
\end{align}\label{transfunctns2}%
\end{subequations}
which brings us from one frame to the other. Such an $SL(2,\C)$ gauge transformation exists provided the matching conditions \eref{matchcond} are satisfied, which has been shown in \eref{transfunctns} above. The spinors are gauge equivalent, but the connections may not: There is, in general, a non-vanishing difference tensor $\ou{[C^{ij}]}{A}{Ba}$ between the two $SL(2,\C)$ connections, and we define it as follows:
\begin{equation}
A^{ij}=h^{ji}\di h^{ij}+h^{ji}A^{ji}h^{ij}+h^{ji}C^{ji}h^{ij}.\label{difftensordef}
\end{equation}

\paragraph{Construction of the action} The action will consists of a contribution from every four-dimensional cell $\mathcal{M}_i$, a boundary term from every interjacent null surface, and a corner term from any two such null surfaces intersecting in a two-dimensional face. The bulk contribution imposes $\Lambda$-flatness \eref{Lflat} of the connection. We allow for the presence of an Immirzi parameter $\beta$ (as in equation \eref{selfdualactn} above), which amounts to work with the \emph{twisted} bulk momentum
\begin{equation}
\Pi_{AB}=\frac{\I}{16\pi\beta G}(\beta+\I)\Sigma_{AB},\label{twsting}
\end{equation}
which is an $\mathfrak{sl}(2,\C)$-valued two-form with dimensions of $\hbar$. The bulk contribution to the action is therefore nothing but the integral
\begin{equation}
\int_{\mathcal{M}_i}\left[\Pi_{AB}\wedge F^{AB}+\frac{8\pi\I\beta\Lambda G}{3}\frac{1}{\beta+\I}\Pi_{AB}\wedge\Pi^{AB}\right]+\CC\label{LBFactn}
\end{equation}
Clearly, variation with respect to the self-dual two-form $\Pi_{AB}$ imposes the desired constraint \eref{Lflat}, i.e. $F_{AB}\propto\Pi_{AB}$, and the variation with respect to the connection implies the integrability condition
\begin{equation}
D\Pi_{AB}=\di\Pi_{AB}-2\ou{A}{C}{(A}\wedge \Pi_{B)C}=0.\label{torsless}
\end{equation}

The next term to add is the three-dimensional boundary term from the interface between two adjacent regions. This boundary term has a twofold job: It cancels the connection variation from the bulk, and it imposes that the intrinsic boundary geometry is null. 
It consists of the covariant symplectic\footnote{Notice: $D^{ij}=\di+[A^{ij},\cdot]$ is the covariant derivative with respect to the $SL(2,\C)$ connection $A^{ij}$, which is the pullback of the self-dual connection from $\mathcal{M}_i$ to $\mathcal{N}_{ij}\subset\partial \mathcal{M}_i$.} potential
\begin{equation}
\int_{\mathcal{N}_{ij}}\Big[\bold{\pi}^{ij}_A\wedge D^{ij}\ell^A_{ij}-\bold{\pi}^{ji}_A\wedge D^{ji}\ell^A_{ji}\Big]+\CC,\label{spinactn}
\end{equation}
plus additional constraints: The glueing conditions \eref{matchcond}, which match the spinors from the two sides and the reality conditions \eref{realcond}, which impose that the area two-from $\boldsymbol{\varepsilon}$ is real. We thus introduce Lagrange multipliers $\omega$, $\Psi_{ab}$ and $\lambda$ and add the terms
\begin{align}\nonumber
\int_{\mathcal{N}_{ij}}\bigg[\omega^{ij}\wedge\big(\bold{\pi}^{ij}_A\ell^A_{ij}-\bold{\pi}^{ji}_A\ell^A_{ji}\big)+
\Psi_{ab}^{ij}\left(\bold{\pi}^{ij}_A{}^a\bold{\pi}^{A}_{ij}{}^a-\bold{\pi}^{ji}_A{}^a\bold{\pi}^{A}_{ji}{}^a\right)+\CC\bigg]+\\
+\int_{\mathcal{N}_{ij}}\lambda^{ij}\wedge\left[\frac{\I}{\beta+\I}\left(\bold{\pi}^{ij}_A\ell^A_{ji}+\bold{\pi}^{ij}_A\ell^A_{ji}\right)+\CC\right]\label{consterms}
\end{align}
to the action. Finally, we also need a corner term to cancel the boundary term $\int_{\partial\mathcal{N}}\bold{\pi}_A\delta \ell^A$ arising from the $\ell^A$-variation on $\mathcal{N}$. 

We sum the bulk action \eref{LBFactn} with the boundary action for the spinors \eref{spinactn} and the constraints \eref{consterms} and also add the right corner term, which we will discuss below. The resulting action is then given by the expression
\begin{align}\nonumber
&S[\Pi,A|\bold{\pi},\ell|\omega,\lambda,\Psi,\alpha]=\\\nonumber
&=2\sum_{i}\int_{\mathcal{M}_i}\left[\Pi_{AB}\, F^{AB}+\frac{8\pi\I\beta\Lambda G}{3}\frac{1}{\beta+\I}\Pi_{AB}\,\Pi^{AB}\right]+\\\nonumber
&+\sum_{\langle ij\rangle}\int_{\mathcal{N}_{ij}}\!\left[\bold{\pi}_{A} (D-\omega)\ell^A-\frac{\lambda}{2}\Big(\frac{\I}{\beta+\I}\bold{\pi}_A\ell^A+\CC\Big)
-\frac{1}{2}{\Psi}_{ab}\uo{\boldsymbol{\pi}}{A}{a}{\boldsymbol{\pi}}^{Ab}\right]+\nonumber\\
&+\sum_{[ijmn]}\int_{\mathcal{C}_{ij}^{mn}}\!\!\!\!\!\!\alpha\,
\big(\ell^{im}_A\,\ell_{in}^A-\ell^{jm}_A\,\ell_{jn}^A+\ell_A^{mj}\ell^A_{mi}-\ell_A^{nj}\ell^A_{ni}\big)+\CC\label{fullactn}
\end{align}
The first sum $\sum_{i}$ goes over all four-dimensional bulk regions $\mathcal{M}_i\in\{\mathcal{M}_1,$ $\mathcal{M}_2,\dots\}$, the second sum goes over all ordered pairs $(\mathcal{M}_i,\mathcal{M}_j)$, which share an interface $\mathcal{N}_{ij}$. This sum crucially contains both possible orientations, i.e. $\sum_{\langle ij\rangle}\int_{\mathcal{N}_{ij}}\dots=\int_{\mathcal{N}_{12}}\dots+\int_{\mathcal{N}_{21}}\dots+\dots$. The last integral is the corner term, which is a sum over all quadruples $(\mathcal{M}_i,\mathcal{M}_j,\mathcal{M}_m,\mathcal{M}_n)$ that share a two-dimensional corner such that $(\mathcal{M}_i,\mathcal{M}_m)$, $(\mathcal{M}_i,\mathcal{M}_n)$ and $(\mathcal{M}_j,\mathcal{M}_m)$ and $(\mathcal{M}_j,\mathcal{M}_n)$ each share a three-dimensional interface (e.g. $\mathcal{M}_i\cap\mathcal{M}_m=\mathcal{N}_{im}$), whereas $(\mathcal{M}_i,\mathcal{M}_j)$ and $(\mathcal{M}_m,\mathcal{M}_n)$ only meet in the corner itself: $\mathcal{M}_i\cap\mathcal{M}_j=\mathcal{M}_m\cap\mathcal{M}_n=\mathcal{C}_{ij}^{mn}$. See \hyperref[fig1]{figure 1} for an illustration. We write $\sum_{[ijmn]}$ to say that any such corner appears with only one possible orientation in the sum. This orientation is chosen arbitrarily and can be absorbed into the definition of $\alpha$, which is a complex-valued two-form at the corner. The variation with respect to $\alpha$ turns out to vanish by taking into account the boundary conditions, which are obtained from the two-dimensional remainder 
\begin{equation}
\int_{\mathcal{C}^{mn}_{ij}}(\bold{\pi}^{in}_A\delta\ell^A_{in}\pm\alpha\ell_A^{im}\delta\ell^A_{in})
\end{equation}
arising from the $\ell^A$-variation of the coupled boundary plus corner terms. The resulting boundary conditions are
\begin{equation}
\varphi^\ast_{\mathcal{C}^{mn}_{ij}}\bold{\pi}_A^{in}=\mp\alpha\ell_A^{im},\label{boundcond}
\end{equation}
with $\varphi^\ast_{\mathcal{C}^{mn}_{ij}}$ denoting the pull-back to the two-dimensional corner and the relative sign depending on the orientation of the corner $\mathcal{C}^{mn}_{ij}$ relative to the null surface $\mathcal{N}_{in}$.

Let us summarise and  briefly explain the role of each term in the action \eref{fullactn}. The first line is the bulk action. It imposes with \eref{twsting} $\Lambda$-flatness \eref{Lflat} of the self-dual connection. Next, there are the integrals over the internal boundaries $\mathcal{N}_{ij}$. The variation of the connection in the bulk yields a remainder at the boundary. The variation of the boundary connection couples then the boundary with the bulk, yielding the constraint
\begin{equation}
\varphi^\ast_{ij}\Pi_{AB}=\frac{1}{2}\bold{\pi}_{(A}^{ij}\ell^{ij}_{B)},
\end{equation}
with $\varphi^{ij}:\mathcal{N}_{ij}\hookrightarrow\mathcal{M}$ denoting the canonical embedding. Next, there is the reality condition: The Lagrange multiplier $\lambda$ is a real-valued one-form on $\mathcal{N}_{ij}$, its variation imposes the reality condition
\begin{equation}
\boldsymbol{\varepsilon}_{ab}=\bar{\boldsymbol{\varepsilon}}_{ab},
\end{equation}
for the area two-form \eref{areaform} as in equation \eref{realcond} above. Finally, there are the glueing conditions \eref{matchcond}, which follow from the variation of the Lagrange multipliers $\omega$ and $\Psi_{ab}=-\Psi_{ba}$, which are continuous across the interface:
\begin{equation}\omega_{ij}=
\omega|_{\mathcal{N}_{ij}}=\omega|_{\mathcal{N}_{ji}},\quad
\Psi_{ab}^{ij}={\Psi}_{ab}|_{\mathcal{N}_{ij}}={\Psi}_{ab}|_{\mathcal{N}_{ji}}.\label{continty1}
\end{equation}
We can then always shift $\omega_{ij}$ by a term proportional to $\lambda_{ij}+\lambda_{ji}$. This allows us to restrict the Lagrange multiplier $\lambda$ such that its sign flips across the interface, in other words 
\begin{equation}
\lambda_{ij}=\lambda|_{\mathcal{N}_{ij}}=-\lambda|_{\mathcal{N}_{ji}}.\label{continty2}
\end{equation}
Notice also that we have used a condensed notation in \eref{fullactn}, we dropped all wedge products, and suppressed all $(ij)$-indices in the boundary variables. The second line in \eref{fullactn} has to be understood, therefore, in the following sense
\begin{equation}
\int_{\mathcal{N}_{ij}}\bold{\pi}_A(D-\omega)\ell^A+\dots\equiv\int_{\mathcal{N}_{ij}}\bold{\pi}_A^{ij}\wedge(D^{ij}-\omega^{ij})\ell^B_{ij}+\dots
\end{equation}
Finally, a word on the Lagrange multipliers: The continuity conditions \eref{continty1} and \eref{continty2} for the Lagrange multipliers $\Psi_{ab}$, $\omega$ and $\lambda$ are part of the definition of the action \eref{fullactn}. The Lagrange multiplier $\omega^{ij}=\omega^{ji}$ is a $U(1)_\C$ connection along the internal boundaries, $\lambda^{ij}=-\lambda^{ji}$ is a real-valued one-form, and $\Psi^{ij}_{ab}=-\Psi^{ij}_{ba}=\Psi^{ji}_{ab}$ is a complex-valued tensor density of weight minus one. The density weight of $\Psi_{ab}$ can be inferred from the definition of the momentum density
\begin{equation}
\uo{\boldsymbol{\pi}}{A}{a}:=\frac{1}{2}\oepsilon^{abc}\bold{\pi}_{Abc},
\end{equation}
which is a spinor-valued vector density of weight one, with  $\oepsilon^{abc}$ denoting the
 metric independent Levi-Civita density $\oepsilon^{abc}$. 

\subsection{Equations of motion}
In this section, we study the equations of motion as derived from the action \eref{fullactn}. This is a preparation for the next section, where we will find a family of explicit solutions representing plane fronted gravitational waves, which are exact solutions of both the discretised theory and general relativity as well. 

\paragraph{} Some of the equations of motion derived from the action \eref{fullactn} have already been mentioned. The variation of the self-dual two-form $\Pi_{AB}$ in the bulk yields the flatness constraint \eref{Lflat}, i.e. 
\begin{equation}
\forall p\in\mathcal{M}_i:F_{AB}(p)=\frac{16\pi\beta\,\Lambda G}{3\I}\frac{1}{\beta+\I}\Pi_{AB}(p).\label{Fsourced}
\end{equation}
We then also have the variation of the self-dual connection $\ou{A}{A}{B}$, which gives the integrability condition $D\Pi_{AB}=0$. This is essentially the torsionless condition. 
Next, there are the internal null boundaries $\{\mathcal{N}_{ij}\}$, where we find a number of additional constraints. The \emph{reality conditions} \eref{realcond} impose that the area two-form $\bold{\varepsilon}=-8\pi\beta G/(\beta+\I)\,\bold{\pi}_A\ell^A$ is real, and they are obtained from the stationary points of the action \eref{fullactn} with respect to variations of the Lagrange multiplier $\lambda$. In the same way, the Lagrange multipliers $\omega$ and $\Psi_{ab}$ are added to the action in order to impose the glueing conditions across the interface, namely: the \emph{area-matching condition} \eref{areamatch} and the \emph{shape-matching} condition \eref{shapematch}. Finally, we have the relation between the bulk and the boundary, which is provided by the glueing condition
\begin{equation}
\varphi^\ast \Pi_{AB}=\frac{1}{2}\bold{\pi}_{(A}\ell_{B)},\label{Psourced}
\end{equation}
which couples the boundary spinors $\bold{\pi}$, $\ell^A$ to the pull-back of the self-dual two-form $\Pi_{AB}$ to the boundary. As we have seen previously in section \ref{sec2.1} above, this is the same as to say that the boundary is null.

The action \eref{fullactn} contains the boundary spinors as additional configuration variables, and the action is stationary with respect to them provided additional equations of motion are satisfied along the system of interfaces. For simplicity, consider only a single such interface $\mathcal{N}$, bounding the four-dimensional regions $\mathcal{M}$ and $\utilde{\mathcal{M}}$, with $\mathcal{N}\subset\partial\mathcal{M}$. This interface will carry spinors  $(\bold{\pi}_A,\ell^A)$ and $(\utilde{\bold{\pi}}_A,\utilde{\ell}^A)$, which represent the induced three-geometry from, say, below and above the interface: If $\varphi:\mathcal{M}\hookrightarrow\mathcal{N}$ (and $\utilde{\varphi}:\utilde{\mathcal{M}}\hookrightarrow\mathcal{N}^{-1}$) is the canonical embedding of the boundary into the bulk, we will have ${\varphi}^\ast\Pi_{AB}={\bold{\pi}}_{(A}{\ell}_{A)}$ and $\utilde{\varphi}^\ast\Pi_{AB}=\utilde{\bold{\pi}}_{(A}\,\utilde{\ell}_{\,B)}$ respectively.

Going back to the definition of the action \eref{fullactn}, we then see that the variation of the boundary spinors yields the equations of motion
\begin{subequations}
\begin{align}
D_a\ell^A&=\left(\omega_a+\frac{\I}{\beta+\I}\lambda_a\right)\ell^A+\Psi_{ab}\uo{\bold{\pi}}{}{Ab},\label{elldiff}\\
D\bold{\pi}_A\,&=-\omega\wedge\bold{\pi}_A-\frac{\I}{\beta+\I}\lambda\wedge\bold{\pi}_A,
\end{align}
and
\begin{align}
&\utilde{D}_a\,\utilde{\ell}^A=\left(\omega_a-\frac{\I}{\beta+\I}\lambda_a\right)\utilde{\ell}^A+\Psi_{ab}\uo{\utilde{\bold{\pi}}}{}{Ab},\label{elltildediff}\\
&\utilde{D}\utilde{\bold{\pi}}_A=-\omega\wedge\utilde{\bold{\pi}}_A+\frac{\I}{\beta+\I}\lambda\wedge\utilde{\bold{\pi}}_A,
\end{align}\label{spinkoeff}%
\end{subequations}
where $D_a$ (and $\utilde{D}_a$) is the covariant derivative with respect to the $SL(2,\C)$ connection $\ou{A}{A}{Ba}=[{\varphi}^\ast\ou{A}{A}{B}]_a $ (and $\ou{\utilde{A}}{A}{Ba}=[\utilde{\varphi}^\ast\ou{A}{A}{B}]_a $), which is the pull-back of the bulk connection in $\mathcal{M}$ (and $\utilde{\mathcal{M}}$) to the interface $\mathcal{N}$ between $\mathcal{M}$ and $\utilde{\mathcal{M}}$. 

\paragraph{Three immediate observations: Integrability, geodesity and the expansion of the null surface} Before we proceed, we need to develop some better intuition for this system of equations \eref{spinkoeff}. First of all, we can see that they are consistent with the  torsionelss condition $D\Pi_{AB}=0$ for the self-dual two-form $\Pi_{AB}$. Indeed, $\varphi^\ast(D\Pi_{AB})=D\varphi^\ast(\Pi_{AB})=D(\bold{\pi}_{(A}\ell_{B)})=\uo{\bold{\pi}}{(A}{a}\uo{\bold{\pi}}{B)}{b}\Psi_{ab}=0$, which vanishes, since the Lagrange multiplier $\Psi_{ab}=-\Psi_{ba}$ is an anti-symmetric tensor density, whereas $\uo{\bold{\pi}}{(A}{a}\uo{\bold{\pi}}{B)}{b}$ is symmetric in $a$ and $b$. We can then, however, also form the anti-symmetric tensor density $\uo{\bold{\pi}}{A}{a}\bold{\pi}^{Ab}=-\uo{\bold{\pi}}{A}{b}\bold{\pi}^{Aa}$ and contract it with $\Psi_{ab}$. The resulting density does not vanish in general and it has an important geometrical interpretation. It measures the expansion $\vartheta_{(\ell)}$ of the null surface. This can be seen as follows. The expansion $\vartheta_{(\ell)}$ can be defined by
\begin{equation}
\di\bold{\varepsilon}=-\vartheta_{(\ell)}\,k\wedge\bold{\varepsilon}\in \Omega^2(\mathcal{N}:\R),
\end{equation}
with $\bold{\varepsilon}$ denoting the canonical two-dimensional area element \eref{arint} on $\mathcal{N}$. The definition of $\vartheta_{(\ell)}$ is gauge dependent\,---\,it depends on a representative $\ell^a$ of the equivalence class of null generators of $\mathcal{N}$, with $k_a$ denoting a dual one-form $k_a\in T^\ast\mathcal{N}:k_a\ell^a=-1$, which is intrinsic to $\mathcal{N}$. In terms of the boundary spinors, the two-dimensional area element is
\begin{equation}
\bold{\varepsilon}=-\frac{8\pi\beta G}{\beta+\I}\bold{\pi}_A\ell^A.\label{artwoformdef}
\end{equation}
Going back to the equations of motion \eref{spinkoeff}, we find
\begin{equation}
\di\bold{\varepsilon}=D\bold{\varepsilon}=-\frac{8\pi\beta G}{\beta+\I}\Psi_{ab}\uo{\bold{\pi}}{A}{a}\uo{\bold{\pi}}{}{Ab}\equiv-\vartheta_{(\ell)} k\wedge\bold{\varepsilon},
\end{equation}
which shows that the density $\Psi_{ab}\uo{\bold{\pi}}{A}{a}\uo{\bold{\pi}}{}{Ab}$ is a measure for the expansion $\vartheta_{(\ell)}$ of the null surface $\mathcal{N}$.

Finally, let us turn to the null generator $\ell^a$ itself. It is geodesic, and this can be seen as follows. First of all, we write this vector field (modulo an overall normalisation) in terms of the boundary spinors, obtaining
\begin{equation}
\ell^a\propto (\beta+\I)\uo{\bold{\pi}}{A}{a}\ell^A,
\end{equation}
which follows from the area-two from as written in terms of the boundary spinors, and $\oepsilon^{abc}\bold{\varepsilon}_{bc}\propto \ell^a$ as in \eref{elldensity}. If we contract this vector field with the soldering form, we get back $\I\ell^A\bar{\ell}^{\bar A}$, see for instance equation \eref{pulltetra} above. Now, the covariant derivative of $\ell^A$ along the null generators is proportional to $\ell^A$ itself, which follows from
\begin{equation}
\ell^aD_a\ell^A\propto\ell^A\Leftrightarrow \ell_A\ell^aD_a\ell^A=0,
\end{equation}
and equation \eref{elldiff} by noting that
\begin{equation}
\ell_A\ell^aD_a\ell^A=\ell^a\Psi_{ab}\uo{\bold{\pi}}{A}{b}\ell^A\propto\Psi_{ab}\ell^a\ell^b=0,
\end{equation}
which is zero since $\Psi_{ab}$ is anti-symmetric. We have thus shown that
\begin{equation}
\ell^bD_b\left(\I\ell^A\bar{\ell}^{\bar A}\right)\propto\I\ell^A\bar{\ell}^{\bar A},
\end{equation}
which means that the integral curves of $\ell^a\equiv\I\ell^A\bar{\ell}^{\bar A}$ are auto-parallel curves with respect to the $SL(2,\C)$ connection. The $SL(2,\C)$ connection is torsionless \eref{torsless}, hence the null generators of $\mathcal{N}$ are geodesics.
\paragraph{Difference tensor} The null surface $\mathcal{N}$ bounds two bulk regions $\mathcal{M}$ (from above) and $\utilde{\mathcal{M}}$ (from below). There are then two $SL(2,\C)$ connections on $\mathcal{N}$, one (namely $\ou{A}{A}{Ba}$) from below the interface, the other (namely $\ou{\utilde{A}}{A}{Ba}$) from above. Their relative strength is given by a difference tensor $\ou{C}{A}{Ba}$, whose algebraic form is determined as follows. First of all, we know that the glueing conditions \eref{matchcond} imply that the spinors $(\bold{\pi}_A,\ell^A)$ and $(\utilde{\bold{\pi}}_A,\utilde{\ell}^A)$ are gauge equivalent, which means that they are related by an $SL(2,\C)$ gauge transformation. We can thus write \begin{equation}
\utilde{\ell}^A\approx\ell^A,\quad\utilde{\bold{\pi}}_A\approx\bold{\pi}_A,\label{gaugeequi}
\end{equation}
where the symbol \qq{$\approx$}, means equality up to gauge transformation as in (\ref{transfunctns}, \ref{transfunctns2}) above. We then have the difference tensor \eref{difftensordef} on $\mathcal{N}$, which is defined as
\begin{equation}
\ou{\utilde{A}}{A}{Ba}\approx\ou{A}{A}{Ba}+\ou{C}{A}{Ba}.\label{difftendef}
\end{equation}
Subtracting the covariant differential of $\ell^A$ from the differential of $\utilde{\ell}^A$, i.e. subtracting \eref{elltildediff} from \eref{elldiff}, and equally $D\bold{\pi}_A$ from $\utilde{D}\utilde{\bold{\pi}}_A$ we then find the conditions
\begin{subalign}
\ou{C}{A}{Ba}\ell^B=&-\frac{2\I}{\beta+\I}\lambda_a\ell^A,\\
\ou{C}{A}{B}\wedge\bold{\pi}^B=&+\frac{2\I}{\beta+\I}\lambda\wedge\bold{\pi}^A.\label{difftens2}
\end{subalign}
This implies that $\ou{C}{A}{Ba}$ admits the decomposition
\begin{equation}
C_{ABa}=-\frac{4\I}{\beta+\I}\ell_{(A}k_{B)}\lambda_a-\frac{4}{\beta+\I}\ell_A\ell_B\Gamma_a,\label{compC}
\end{equation}
where the component function $\Gamma_a$ (a complex-valued one-form on $\mathcal{N}$) is subject to the algebraic constraint
\begin{equation}
\lambda\wedge\bold{\mu}=\Gamma\wedge\bold{\varepsilon},\label{Gammalambdacons}
\end{equation}
which is a consequence of \eref{difftens2}. 
Finally, the difference tensor is subject to one additional constraint: The field strengths as induced from the two sides are gauge equivalent, which is an immediate consequence of $F_{AB}$ being sourced by $\Pi_{AB}$ (as in \ref{Fsourced}), and $\varphi^\ast\Pi_{AB}$ being sourced by the boundary spinors (as in \ref{Psourced}). The boundary spinors are gauge equivalent (see \eref{gaugeequi}), which implies (on $\mathcal{N}$) that $\utilde{F}_{AB}\approx F_{AB}$, which is the same as to say
\begin{equation}
D\ou{C}{A}{B}+\ou{C}{A}{C}\wedge\ou{C}{C}{B}=0.\label{exdiffdiff}
\end{equation}
In the next section, we will demonstrate that explicit solutions to these equations exist: Plane gravitational waves solve the system of equations (\ref{spinkoeff}, \ref{compC}, \ref{exdiffdiff}, \ref{Gammalambdacons}) for certain boundary spinors and a definite difference tensor $\ou{C}{A}{B}$ at the interface.

\subsection{Special solutions: Plane-fronted gravitational waves}\label{sec3.4}
This section is dedicated to finding explicit solutions to the equations of motion in the neighbourhood of an interface. Rather than exploring the entire solution space, we study only a single family of solutions, thus giving a constructive proof of existence: There are non-trivial\footnote{We will find solutions with a non-vanishing distributional Weyl tensor at the interface.} solutions to the equations of motion derived from the action \eref{fullactn}, and the particular solutions thus constructed are distributional solutions of Einstein's equations as well.

Consider thus a single interface $\mathcal{N}$, bounding the four-dimensional regions $\mathcal{M}$ and $\utilde{\mathcal{M}}$ from above and below. We set the cosmological constant to zero, hence $\mathcal{M}$ and $\utilde{\mathcal{M}}$ are flat, and we assume, in addition, that the lightlike interface $\mathcal{N}$ is the flat hyperplane $x^0=x^3$, where $\{x^\mu\}$ are inertial coordinates in $\mathcal{M}$. The glueing conditions \eref{metrmatch} imply that the intrinsic geometry of $\mathcal{N}$ is the same from the two sides. The discontinuity in the metric can, therefore, only be in the transversal direction, which motivates the following ansatz for the tetrad across the interface
\begin{align}\nonumber
e^\alpha=\ell^\alpha \di u+&\left[k^\alpha+
\Theta(v)\left(fm^\alpha+\bar{f}\bar{m}^\alpha+f\bar{f}\ell^\alpha\right)\right]\di v+\\
&\;+\bar{m}^\alpha\di z+m^\alpha\di\bar{z}.\label{ansatz}
\end{align}
The vectors $\{\ell^\alpha,k^\alpha,m^\alpha,\bar{m}^\alpha\}$ are a normalised null tetrad: $\bar{m}_\alpha m^\alpha=-\ell^\alpha k_\alpha=1$, and they are all  constant with respect to the ordinary derivative\footnote{This is the covariant derivative for $v<0$.} $\di\ell^\alpha=\di k^\alpha=\di m^\alpha=0$. The coordinate functions are $u=(x^0+x^3)/\sqrt{2}$, $v=(x^0-x^3)/\sqrt{2}$ and $z=(x^1+\I x^2)/\sqrt{2}$, while $\Theta(v)$ denotes the Heaviside step function. The only freedom is in $f$, which is a complex-valued function of $z$ and $u$ alone: $f\equiv f(u,z,\bar{z})$.  The ansatz \eref{ansatz} describes a class of impulsive gravitational wave solutions widely known in the literature \cite{Griffiths:1991zp}, it includes plane fronted gravitational waves, which may describe e.g.\ the gravitational field of a massless particle \cite{Aichelburg1971}.

	We now have to make sure that our ansatz \eref{ansatz} is compatible with the equations of motion as derived from the action \eref{fullactn}. Clearly, the equations of motion in the bulk, i.e. $F_{\alpha\beta}=0$ and $D\Sigma_{\alpha\beta}=0$ as in \eref{Fsourced} are satisfied for $D=\di$ and $\Sigma_{\alpha\beta}=e_\alpha\wedge e_\beta$ for $v<0$. At the interface, we have a discontinuity in the transversal $v$ direction, which yields a discontinuity in the connection across the interface. The strength of this discontinuity is measured by a difference tensor, which must be matched to \eref{difftendef} and \eref{compC}. We thus have to compute the spin connection and take the difference between the two sides. The spin connection is the unique solution to the torsionless condition
\begin{equation}
De^\alpha=\di e^\alpha+\ou{A}{\alpha}{\beta}\wedge e^\beta=0.
\end{equation}
The unique solution for $\ou{A}{\alpha}{\beta}$ in terms of $f(u,z,\bar{z})$ gives a rather lengthy expression. For the moment, we are only interested in the pull-back  to the hyperplane $v=0$, which defines the difference tensor
\begin{align}
\ou{\Delta}{\alpha}{\beta}:=\lim_{v\searrow 0}\Big(\underleftarrow{\ou{A}{\alpha}{\beta}}\Big|_{v}-\underleftarrow{\ou{A}{\alpha}{\beta}}\Big|_{-v}\Big)\label{Deltadeff}
,
\end{align}
where $\ou{\underleftarrow{A}}{\alpha}{\beta}\big|_v$ is the pull-back of the spin connection to a $v=\mathrm{const}.$ surface. To compare this equation \eref{Deltadeff} with the definition \eref{difftendef} of the difference tensor $\ou{C}{A}{Ba}$, we decompose $\ou{\Delta}{\alpha}{\beta a}$ into its self-dual and anti-self-dual components. Following the conventions of \eref{selfdualcompon}, we call them $\ou{\Delta}{A}{B a}$ and $\ou{\bar\Delta}{\bar A}{\bar B a}$, with $\ou{\Delta}{A\bar A}{B\bar Ba}=\delta^{\bar A}_{\bar B}\ou{\Delta}{A}{Ba}+\delta^{A}_{B}\ou{\bar\Delta}{\bar A}{\bar Ba}$. We then introduce a normalised spin dyad $\{k^A,\ell^A\}:k_A\ell^A=1$, which is covariantly constant in $\mathcal{M}$, i.e.\ $\di\ell^A=0=\di k^A$, and which is related to the null tetrad $\{\ell^\alpha,k^\alpha,m^\alpha,\bar{m}^\alpha\}$ in the canonical way
\begin{equation}\left.
\begin{split}
\ell^\alpha=-\frac{1}{\sqrt{2}}\uo{\sigma}{A\bar A}{\alpha}\ell^A\bar{\ell}^{\bar A},\;\quad 
&m^\alpha=-\frac{1}{\sqrt{2}}\uo{\sigma}{A\bar A}{\alpha}\ell^A\bar{k}^{\bar A},\\
k^\alpha=-\frac{1}{\sqrt{2}}\uo{\sigma}{A\bar A}{\alpha}k^A\bar{k}^{\bar A},\quad 
&\bar{m}^\alpha=-\frac{1}{\sqrt{2}}\uo{\sigma}{A\bar A}{\alpha}k^A\bar{\ell}^{\bar A}.
\end{split}\quad\right\}\label{nulltetra}
\end{equation}
A straightforward and rather lengthy calculation gives then the difference tensor in terms of the triadic basis $\{\ell_A\ell_B,k_Ak_B,k_{(A}\ell_{B)}\}$. The result is
\begin{align}\nonumber
\Delta_{AB}=&-\frac{1}{2}\ell_{(A}k_{B)}\left(\di z\,\partial_uf+\di\bar{z}\,\partial_u\bar{f}\right)+\\
&+\frac{1}{2}\ell_A\ell_B\left(\di u\,\partial_u f+2\di z\,\partial_z f+\di\bar{z}\,\partial_{\bar{z}}f+\di\bar{z}\partial_z\bar{f}\right).
\end{align}
In order to satisfy the equations of motion, this difference tensor must be equal to $\ou{C}{A}{Ba}$ as derived from the equations of motion for the action \eref{fullactn}. Going back to the decomposition of $\ou{C}{A}{Ba}$ into $\ell^A$ and $k^A$, as in \eref{compC} above, we thus find the conditions
\begin{subalign}
-\frac{8\I}{\beta+\I}\lambda&=\di z\,\partial_u f+\di\bar{z}\,\partial_u \bar{f},\label{conncoeff1}\\
+\frac{8}{\beta+\I}\Gamma&=\partial_uf \di u+2\partial_z f\di z+\left(\partial_{\bar z}f+\partial_z\bar{f}\right)\di\bar{z}.\label{conncoeff2}
\end{subalign}
Now, the Lagrange multiplier $\lambda$ is real (it imposes the reality conditions \eref{realcond}), while $\Gamma$ is complex. Unless the Barbero\,--\,Immirzi parameter goes to zero, which is a singular limit in the original selfdual action \eref{selfdualactn}, the equation \eref{conncoeff1} can be only imposed, therefore, for $\lambda=0$. This implies, in turn,
\begin{equation}
\partial_u f=0.
\end{equation}
Going back to our ansatz \eref{ansatz} for the tetrad, we thus see that the null vector $\partial_u^a=\ell^a$ is a Killing vector. Having solved the equations of motion for $v<0$, and  matched the geometries across the interface, we are now left to solve the equations of motion for $v>0$. The only missing condition is to impose that the curvature vanishes for $v>0$. A straightforward calculation reveals that this is possible if and only if
\begin{equation}
\partial_{\bar z} f-\partial_z\bar{f}=0.
\end{equation}

It is also instructive to have a look at the boundary spinors and compute them explicitly. Going back to our initial ansatz \eref{ansatz} for the tetrad and taking also into account the parametrisation \eref{nulltetra} of the null tetrad, we immediately see that we are in a gauge for which $\ell^A=\utilde{\ell}^A$. For $\bold{\pi}_A$ and $\utilde{\bold{\pi}}_A$, on the other hand, we need to consider the Pleba\'nski two-form $\Sigma_{\alpha\beta}=e_\alpha\wedge e_\beta$, and determine its self-dual part as in \eref{selfdualforms}. The result of this straightforward exercise returns the momentum spinors
\begin{equation}
\bold{\pi}_A=\utilde{\bold{\pi}}_A=\frac{\I}{8\pi\beta G}(\beta+\I)\left[\ell_A\di\bar{z}\wedge \di u+k_A\di\bar{z}\wedge \di z\right].
\end{equation}
The two-dimensional volume element on $\mathcal{N}$, as given by equation \eref{arint}, is then simply
\begin{equation}
\bold{\varepsilon}=-\I\di\bar{z}\wedge\di z=\di x^1\wedge \di x^2.
\end{equation}
Going back to the equations of motion for the spinors, i.e. equations \eref{spinkoeff} above, we can then also determine the missing Lagrange multipliers: We get $\Psi_{ab}=0$ and $\omega_a=0$ and $\lambda_a=0$. This fully determines the geometry in terms of the boundary  variables.

Finally, we can now also compute the distributional curvature tensor across the interface. Following Penrose's conventions \cite{penroserindler}, we compute the irreducible components of the field strength of the $SL(2,\C)$ connection, namely
\begin{equation}
F_{ABcd}\equiv F_{ABC\bar C D\bar D}=-\bar{\epsilon}_{\bar C\bar D}\Psi_{ABCD}-\epsilon_{CD}\Phi_{AB\bar C\bar D},
\end{equation}
where the Weyl spinor $\Psi_{ABCD}$ represents the irreducible spin $(2,0)$ component, and $\Phi_{AB\bar A\bar B}$ is the traceless part of the Ricci tensor, which is the irreducible spin $(1,1)$ component of the Riemann curvature tensor. Both components vanish everywhere except at the null surface $v=0$, where there occurs a distributional curvature singularity. The Weyl spinor is
\begin{equation}
\Psi_{ABCD}=\delta(v)\,\partial_z f\,\ell_A\ell_B\ell_C\ell_D,\label{Weylspin}
\end{equation}
and the traceless part of the Ricci tensor is determined to be
\begin{equation}
\Phi_{AB\bar A\bar B}=\delta(v)\left(\partial_{\bar z}f+\partial_z \bar{f}\right)\ell_A\ell_B\bar{\ell}_{\bar A}\bar{\ell}_{\bar B}.\label{Riccispin}
\end{equation}
The resulting geometry is a solution to Einstein's equation with a distributional source
\begin{equation}
T_{ab}=\frac{1}{8\pi G}\delta(v)\left(\partial_{\bar z}f+\partial_z \bar{f}\right)\ell_a\ell_b.\label{Ttensor}
\end{equation}
If, in particular, $f$ is holomorphic, i.e. $\partial_{\bar z}f=0$ for all $z$, we have a solution of the vacuum Einstein equations, the other extreme is $f(z)=f_o/(4\pi z)$, which generates the gravitational field of a massless point particle \cite{Aichelburg1971}.

\section{Hamiltonian formulation, gauge symmetries}\label{sec4}
\subsection{Space-time decomposition of the boundary action}
The main purpose of this paper is to open up a new road towards non-perturbative quantum gravity. We have defined the action \eref{fullactn} and demonstrated that explicit solutions exist, which have a  non-vanishing (yet distributional) Weyl curvature \eref{Weylspin} in the neighbourhood of a null surface. The next logical step is to study the Hamiltonian formulation of the theory.

First of all, we note that the bulk action \eref{LBFactn} is topological. All physical degrees of freedom sit, therefore, either at the system of null surfaces $\{\mathcal{N}_{ij}\}$ or at the two-dimensional corners $\{\mathcal{C}_{ij}^{mn}\}$. As long as we are concerned with the canonical analysis on only one such null surface $\mathcal{N}$ alone, we can then also work with a simplified  action, which is found by integrating out the self-dual two-form $\Pi_{AB}$ in the bulk. We insert the equations of motion \eref{Fsourced} for the self-dual two-form back into the bulk action,  thus obtaining
\begin{align}\nonumber
2\int_{\mathcal{M}}\left(\Pi_{AB}\wedge F^{AB}+\frac{1}{\gamma}\Pi_{AB}\wedge\Pi^{AB}\right)\stackrel{\text{EOM}}{=}-\frac{\gamma}{2}\int_{\mathcal{M}}F_{AB}\wedge F^{AB}=\\
=\frac{\gamma}{2}\int_{\partial\mathcal{M}}\operatorname{Tr}\left[A\wedge \di A+\frac{2}{3} A\wedge A\wedge A\right]=\frac{\gamma}{2}S_{\mathrm{CS},\,\partial\mathcal{M}}[A],\label{CSactn}
\end{align}
which is the self-dual Chern\,--\,Simons action with complex-valued coupling constant
\begin{equation}
\gamma=\frac{3}{8\pi\I\,\beta\Lambda G}(\beta+\I).\label{gammadef}
\end{equation}
The appearence of the $SL(2,\C)$ Chern\,--\,Simons action in a four-dimensional theory may come as a surprise, but it has been anticipated by several authors, who have suggested that the so-called Kodama state, which is the exponential of the Chern\,--\,Simons functional for the underlying gauge group, plays a significant role for quantum gravity in four-dimensions \cite{kodama2, kodama1, Smolin:2010iq, Wieland:2011de,  Haggard:2014xoa,Haggard:2015yda}. This paper confirms these early expectations.

The Chern\,--\,Simons action \eref{CSactn} is evaluated over the three-boundary $\partial\mathcal{M}_i$ of the four-dimensional cells $\{\mathcal{M}_1,\mathcal{M}_2,\dots\}$. Each one of them has has the topology of a four ball, whose boundary $\partial\mathcal{M}_i$ is homeomorphic to a three-sphere tessellated into three-dimensional regions $\mathcal{N}_{ij}:\partial\mathcal{M}_i=\bigcup_j\mathcal{N}_{ij}$. For definiteness, consider only one such three-surface $\mathcal{N}$ at the boundary between two bulk regions, say $\mathcal{M}$ and $\utilde{\mathcal{M}}$. Integrating out the self-dual two-form $\Pi_{AB}$ in  $\mathcal{M}$ and $\utilde{\mathcal{M}}$, yields, therefore, two copies of the self-dual $SL(2,\C)$ Chern\,--\,Simons action \eref{CSactn} along the interjacent null surface $\mathcal{N}$. Each one of these $SL(2,\C)$ Chern\,--\,Simons  actions is itself coupled to the respective boundary spinors $(\bold{\pi}_A,\ell^A)$ and $(\utilde{\bold{\pi}}_A,\utilde{\ell}^A)$ from either side. The resulting coupled action is therefore given by
\begin{align}\nonumber
S_{\mathcal{N}}&[A,\utilde{A}|\bold{\pi},\ell,\utilde{\bold{\pi}},\utilde{\ell}|\lambda,\omega,\Psi]=\frac{\gamma}{2}S_{\mathrm{CS},\,\mathcal{N}}[A]-\frac{\gamma}{2}S_{\mathrm{CS},\,\mathcal{N}}[\utilde{A}]+\\
&+\int_{\mathcal{N}}\!\left[\bold{\pi}_{A} (D-\omega)\ell^A-\frac{\lambda}{2}\Big(\frac{\I}{\beta+\I}\bold{\pi}_A\ell^A+\CC\Big)
-\frac{1}{2}{\Psi}_{ab}\uo{\boldsymbol{\pi}}{A}{a}{\boldsymbol{\pi}}^{Ab}\right]+\nonumber\\
&-\int_{\mathcal{N}}\!\left[\utilde{\bold{\pi}}_{A} (\utilde{D}-\omega)\utilde{\ell}^A+\frac{\lambda}{2}\Big(\frac{\I}{\beta+\I}\utilde{\bold{\pi}}_A\utilde{\ell}^A+\CC\Big)
-\frac{1}{2}{\Psi}_{ab}\uo{\utilde{\boldsymbol{\pi}}}{A}{a}{\utilde{\boldsymbol{\pi}}}^{Ab}\right]+\nonumber\\&+\CC\label{bndryactn}
\end{align}
All terms in this boundary action have a straightforward geometrical interpretation: Variation with respect to the self-dual connection imposes that the field strength $F_{AB}$ on $\mathcal{N}$ is sourced\footnote{The same is true for $\utilde{F}_{AB}$ and $-\utilde{\bold{\pi}}_{(A}\utilde{\ell}_{B)}/\gamma$.} by $-\bold{\pi}_{(A}\ell_{B)}/\gamma$, which is consisted with what we have seen before: By \eref{fluxparam} and \eref{pispindef}, this source is itself nothing but the self-dual area two-form $\Sigma_{AB}$ times $\Lambda/3$, hence $F_{AB}=\Lambda/3\,\Sigma_{AB}$ as in pure (anti-)de\,Sitter space. The reality condition \eref{realcond}, which is obtained from the variation of the Lagrange multiplier $\lambda$, imposes then that $\Sigma_{AB}$ is geometric, i.e.\ compatible with the existence of a signature $(0$$+$$+)$ null metric $q_{ab}$ on $\mathcal{N}$. Finally, we have the glueing conditions \eref{matchcond}, which are obtained by demanding that the action be stationary with respect to variations of the multipliers $\omega$ and $\Psi_{ab}$. The glueing conditions impose\footnote{See section \ref{sec3.1} above.} continuity across the interface: The intrinsic geometry of $\mathcal{N}$ is the same whether we compute it from the boundary spinors on either side of the interface.

Next, we write the action in a Hamiltonian form. This requires a clock\,---\,a foliation $\mathcal{N}\simeq[0,1]\times\mathcal{S}$ and a vector field $t^a:t^a\partial_a t=1$, which is transversal to the two-dimensional $t=\mathrm{const}.$ surfaces\footnote{For the moment, we assume them to be closed $\partial\mathcal{S}=\emptyset$. A more complete Hamiltonian analysis taking into account also the corner terms and boundary conditions at the two-dimensional intersections (see \hyperref[fig1]{figure 1}) will be left for the future.} $\mathcal{S}_t=\{t\}\times\mathcal{S}$.  At the level of the action, there is no preferred such vector field $t^a$. This should strike us as a surprise: A three-dimensional null surface always \emph{has} a preferred time direction: the direction of its null generators. So how can it be that our action \eref{bndryactn}, which is meant to be an action for a null surface $\mathcal{N}$, lacks such a preferred structure? The answer is simple: In our theory, there is no metric formulation to begin with, the metric is a derived or composite field. It exists only on-shell\,---\,only if the reality conditions are satisfied. Our fundamental configuration variables are the boundary spinors $\bold{\pi}_A$, $\ell^A$ and $\utilde{\bold{\pi}}_A$, $\utilde{\ell}^A$, and a generic configuration of them does not define a signature $(0$$+$$+$$)$ metric $q_{ab}$ on $\mathcal{N}$. Only for those configurations that satisfy the reality conditions \eref{realcond} can such a metric be defined, according to the construction that has been given in section \ref{sec3.1} above.

We then choose a time function $t$, which is a mere coordinate, and a transversal vector field $t^a:t^a\partial_at=1$ and decompose the configuration variables into space and time components. With a slight abuse of notation, we denote the pull-back of the $SL(2,\C)$ bulk connection $\ou{A}{A}{B}\in\Omega^1(\mathcal{M}:\mathfrak{sl}(2,\C))$ to the $t=\mathrm{const}.$ slices simply by
\begin{equation}
\ou{A}{A}{Ba}=\mathrm{\varphi}_t^\ast[\ou{A}{A}{B}]_a,\label{2dAdef}
\end{equation}
where $\mathrm{\varphi}_t:\mathcal{S}\hookrightarrow\mathcal{N}\subset\mathcal{M},p\mapsto(t,p)$ is the canonical embedding of the $t= \mathrm{const}.$ slices into the three-boundary $\mathcal{N}$. Equation \eref{2dAdef} defines the spatial components of the connection. Its $t$-component\footnote{The hook operator \qq{$\hook$} denotes the interior product of a vector $V$ with a $p$-form $\omega$, i.e. $(V\hook\omega)(X,Y,\dots)=\omega(V,X,Y,\dots)$.} defines the Lagrange multiplier
\begin{equation}
\ou{\Lambda}{A}{B}=\mathrm{\varphi}_t^\ast[t\hook\ou{A}{A}{B}].\label{Lambdadef}
\end{equation}
 Finally, we also have the velocity
\begin{equation}
\ou{\dot{A}}{A}{Ba}=\mathrm{\varphi}_t^\ast[\mathcal{L}_t\ou{A}{A}{B}]_a,
\end{equation}
with $\mathcal{L}_t$ denoting the Lie derivative along the vector field $t^a$.

We repeat the $2+1$ decomposition for the spinor-valued two-form $\bold{\pi}_A$ and write, accordingly
\begin{subequations}
\begin{align}
\sfpi_A&={\varphi}_t^\ast[\bold{\pi}_A]=\frac{1}{2}\oepsilon^{ab}[\varphi^\ast_t\bold{\pi}_A]_{ab},\\
\chi_{Aa}&={\varphi}_t^\ast[t\hook\bold{\pi}_A]_a.\label{chidef}
\end{align}\label{2dspins1}%
\end{subequations}
Finally, we define the velocities
\begin{subalign}
\dot{\ell}^A&=\mathcal{L}_t\ell^A,\\
\dot{\sfpi}_A&=\varphi^\ast_t[\mathcal{L}_t\bold{\pi}_A].
\end{subalign}

In the following, we restrict ourselves to those parts of configuration space, where the area element \eref{arint} is non-degenerate on $\mathcal{S}$, hence we assume
\begin{equation}
{\sfpi}_A\ell^A\neq 0.
\end{equation}
This allows us to use the pair $(\sfpi^A,\ell^A)$ as a basis in $\C^2$, such that we can write the spinor-valued one-form $\ou{\chi}{A}{a}\in\Omega^1(\mathcal{S}:\C^2)$, which is the time-space component of the two-form $\bold{\pi}_A$, in terms of components
\begin{equation}
\chi_{Aa}=\bold{U}_a{\sfpi}_A+V_a\ell_A.
\end{equation}
Notice, the component functions $\bold{U}_a$ and $V_a$ have different density weights: $V_a$ is a one-form on $\mathcal{S}$, while $\bold{U}_a$ is an inverse density one-form, which follows from the fact that ${\sfpi}_A=\varphi^\ast_t\bold{\pi}_A$ is a two-form (hence a density) on $\mathcal{S}$. Both $\bold{U}_a$ and $V_a$ are Lagrange multipliers, since the boundary action \eref{bndryactn} contains no derivatives of them. Yet they are not completely arbitrary: The matching conditions \eref{matchcond} and the reality conditions \eref{realcond} impose constraints on them. 
The reality conditions \eref{realcond}, which follows from the variation of the action \eref{bndryactn} with respect to the Lagrange multiplier $\lambda$, implies
\begin{equation}
\bold{U}_a=\bar{\bold{U}}_a.
\end{equation}
On the other hand, there are the matching conditions \eref{matchcond}, which follow from the variation of the boundary action \eref{bndryactn} with respect to $\Psi_{ab}$ and $\omega$, and they simply yield
\begin{equation}
V_a=\utilde{V}_a,\quad \bold{U}_a=\utilde{\bold{U}}_a,
\end{equation}
where $\utilde{\bold{U}}_a$ and $\utilde{V}_a$ are the components of $\utilde{\chi}_{Aa}$ with respect to the boundary spinors from the other side of the interface, i.e.
\begin{subequations}
\begin{align}
\utilde{\sfpi}_A&=\varphi^\ast_t\utilde{\bold{\pi}}_A=\frac{1}{2}\oepsilon^{ab}\utilde{\bold{\pi}}_{Aab},\\
\ou{\utilde{\chi}}{A}{a}&=\mathrm{\varphi}_t^\ast[t\hook\utilde{\bold{\pi}}_A]_a=\utilde{\bold{U}}_a\utilde{\sfpi}^A+\utilde{V}_a\utilde{\ell}^A.
\end{align}\label{2dspins2}%
\end{subequations}
It is then useful to dualise the component functions $\bold{U}_a$ and $V_a$. We take the canonical Levi-Civita density $\oepsilon^{ab}$ on $\mathcal{S}$, and define
\begin{equation}
N^a:=\oepsilon^{ab}\bold{U}_b,\quad \bold{J}^a:=\oepsilon^{ab}V_b,
\end{equation}
where $N^a\in T\mathcal{S}$ is a tangent vector, while $\bold{J}^a\in\Omega^2(\mathcal{S}:T\mathcal{S})$ is a vector-valued density.

We insert the $2+1$ decompositions for both the connection (i.e. \ref{2dAdef}, \ref{Lambdadef}) and for the boundary spinors (i.e. \ref{2dspins1}, \ref{2dspins2}) back into the boundary action, and get
\begin{align}
S_{\partial\mathcal{N}}&= \nonumber\\
 =&\int\di t\int_{\mathcal{S}}\bigg[\frac{\gamma}{2}\oepsilon^{ab}\left({A}_{ABa}\ou{{\dot{A}}}{AB}{b}-\Lambda^{AB}{F}_{ABab}\right)+{\sfpi}_A\left(\dot{\ell}^A+\ou{\Lambda}{A}{B}\ell^B-\varphi\ell^A\right)\bigg]+\nonumber\\
-&\int\di t\int_{\mathcal{S}}\bigg[\frac{\gamma}{2}\oepsilon^{ab}\left(\utilde{A}_{ABa}\ou{\utilde{\dot{A}}}{AB}{b}-\utilde{\Lambda}^{AB}\utilde{F}_{ABab}\right)\!+\utilde{{\sfpi}}_A\left(\utilde{\dot{\ell}}^A+\ou{\utilde{\Lambda}}{A}{B}\utilde{\ell}^B-\varphi\utilde{\ell}^A\right)\bigg]+\nonumber\\
-&\int\di t\int_{\mathcal{S}}\bigg[\frac{N}{2}\bigg(\frac{\I}{\beta+\I}\big({\sfpi}_A\ell^A+\utilde{{\sfpi}}_A\utilde{\ell}^A\big)+\CC\bigg)+\nonumber\\
&\hspace{1em}+\bold{J}^a\big(\ell_A{D}_a\ell^A-\utilde{\ell}\utilde{{D}}_a\utilde{\ell}^A\big)
+N^a\big({\sfpi}_A{D}_a\ell^A-\utilde{{\sfpi}}_A\utilde{{D}}_a\utilde{\ell}^A\big)\bigg]+\CC,\label{2+1splitactn}
\end{align}
where
\begin{equation}
\varphi=t^a\omega_a,\quad N=t^a\lambda_a
\end{equation}
are the time components of the Lagrange multipliers imposing the area-matching constraint \eref{areamatch}, and the reality conditions \eref{realcond}.

\subsection{Phase space, symplectic structure, constraints}
 Going back to the 2+1 split \eref{2+1splitactn}  of the action, we can immediately read off the symplectic structure. First of all, we see, that the spinors $\ell^A$, $\utilde{\ell}^A$ are canonical conjugate to the densitised momentum spinors $\sfpi_A=\varphi^\ast_t[\bold{\pi}_A]$ and $\utilde{\sfpi}=\varphi^\ast[\utilde{\bold{\pi}}_A]$. The fundamental Poisson brackets are
\begin{subequations}
\begin{align}
\big\{{\sfpi}_A(p),\ell_B(q)\big\}=&+\epsilon_{AB}\delta^{(2)}(p,q),\\
\big\{\utilde{{\sfpi}}_A(p),\utilde{\ell}_B(q)\big\}=&-\epsilon_{AB}\delta^{(2)}(p,q),
\end{align}\label{Poiss1}%
\end{subequations}
where $\delta^{(2)}(p,q)$ is the Dirac delta distribution\,---\,a scalar density\,---\,on $\mathcal{S}$. Next, we have the canonical symplectic structure for the $SL(2,\C)$ connection $\ou{A}{A}{Ba}=\varphi^\ast_t[\ou{A}{A}{B}]_a$ on $\mathcal{S}$, which is given by
\begin{subequations}
\begin{align}
\big\{\ou{A}{AB}{a}(p),A_{CDb}(q)\big\}=&+\frac{1}{\gamma}\uepsilon_{ab}\delta^{(A}_C\delta^{B)}_D\delta^{(2)}(p,q),\\
\big\{\ou{\utilde{A}}{AB}{a}(p),\utilde{A}_{CDb}(q)\big\}=&-\frac{1}{\gamma}\uepsilon_{ab}\delta^{(A}_C\delta^{B)}_D\delta^{(2)}(p,q),
\end{align}\label{Poiss2}%
\end{subequations}
where $\uepsilon_{ab}$ is the inverse Levi-Civita density on $\mathcal{S}$, implicitly defined through $\oepsilon^{ac}\uepsilon_{bc}=\delta^a_b$. Furthermore, $\gamma$ denotes the complex-valued coupling constant \eref{gammadef}, which has dimensions of $\hbar$. Its real and imaginary parts are related by the Barbero\,--\,Immirzi parameter $\beta$. The Poisson brackets \eref{Poiss1} and \eref{Poiss2} and their complex conjugate, e.g. $\{\bar\sfpi_{\bar A}(p),\bar\ell_{\bar B}(q)\}=\bar{\epsilon}_{\bar A\bar B}\delta^{(2)}(p,q)$,  determine the phase-space completely, all other Poisson brackets among the canonical variables vanish.

Next, we have the constraints. The variation with respect to the $\mathfrak{sl}(2,\C)$ Lie algebra elements $\ou{\Lambda}{A}{B}$ and $\ou{\utilde{\Lambda}}{A}{B}$ in the action \eref{2+1splitactn} yields the Gauss constraints
\begin{subequations}
\begin{align}
G_{AB}[\Lambda^{AB}]=\int_{\mathcal{S}}\Lambda^{AB}\left[\frac{\gamma}{2}\oepsilon^{ab}F_{ABab}+\sfpi_A\ell_B\right]\stackrel{!}{=}0,\\
\utilde{G}_{AB}[\utilde{\Lambda}^{AB}]=\int_{\mathcal{S}}\utilde{\Lambda}^{AB}\left[\frac{\gamma}{2}\oepsilon^{ab}\utilde{F}_{ABab}+\utilde{\sfpi}_A\utilde{\ell}_B\right]\stackrel{!}{=}0.
\end{align}\label{Gausscons}%
\end{subequations}%
Variation with respect to $N$ yields the scalar constraint
\begin{equation}
S[N]=\int_{\mathcal{S}}N\left[\frac{\I}{\beta+\I}\big(\sfpi_A\ell^A+\utilde{\sfpi}_A\utilde{\ell}^A\big)+\CC\right]\stackrel{!}{=}0.\label{Skalarcons}
\end{equation}
The Lagrange multiplier $N^a$ gives rise to the vector constraint
\begin{equation}
H_a[N^a]=\int_{\mathcal{S}}N^a\left[\sfpi_AD_a\ell^A-\utilde{\sfpi}_A\utilde{D}_a\utilde{\ell}^A+\CC\right]\stackrel{!}{=}0.\label{Vekcons}
\end{equation}
Finally, we have the matching constraints
\begin{subequations}
\begin{align}
M[\varphi]&=\int_{\mathcal{S}}\varphi\left(\sfpi_A\ell^A-\utilde{\sfpi}_A\utilde{\ell}^A\right)\stackrel{!}{=}0,\label{match1}\\
M_a[\bold{J}^a]&=\int_{\mathcal{S}}\bold{J}^a\left(\ell_AD_a\ell^A-\utilde{\ell}_A\utilde{D}_a\utilde{\ell}^A\right)\stackrel{!}{=}0,\label{match2}
\end{align}%
\end{subequations}%
which are obtained from the variation of the action \eref{2+1splitactn} with respect to the Lagrange multipliers $\varphi$ and $\bold{J}^a$.

Thus far concerning the constraints. The evolution equations, on the other hand, assume a Hamiltonian form as well: For any phase space functional $F$, its time evolution is governed by the Hamilton equations
\begin{equation}
\frac{\di}{\di t}F=\left\{\bold{H},F\right\},\label{Hamflow}
\end{equation}
where the corresponding Hamiltonian is the sum over all constraints of the system
\begin{align}\nonumber
\bold{H}=S[N]+H_a[N^a]+\Big(&G_{AB}[\Lambda^{AB}]+\\
&+\utilde{G}_{AB}[\utilde{\Lambda}^{AB}]+M[\varphi]+M_a[\bold{J}^a]+\CC\Big).\label{canHam}
\end{align}

\subsection{Constraint algebra and gauge symmetries}
Having defined the Hamiltonian, we proceed to calculate the Poisson algebra among the constraints and check whether the constraints are preserved under the Hamiltonian flow \eref{Hamflow}. 

For the Gauss constraints \eref{Gausscons} the situation is straightforward. We recover two copies of local $SL(2,\C)$ gauge transformations: If $\ou{[\Lambda_i]}{A}{B}$ and $\ou{[\utilde{\Lambda}_i]}{A}{B}$ are $\mathfrak{sl}(2,\C)$-valued test functions on $\mathcal{S}$, we find
\begin{subequations}
\begin{align}
\big\{G_{AB}[\Lambda^{AB}_1],G_{CD}[\Lambda^{CD}_2]\big\}&=G_{AB}\left[[\Lambda_1,\Lambda_2]^{AB}\right],\\
\big\{\utilde{G}_{AB}[\utilde{\Lambda}^{AB}_1],\utilde{G}_{CD}[\utilde{\Lambda}^{CD}_2]\big\}&=\utilde{G}_{AB}\left[[\utilde{\Lambda}_1,\utilde{\Lambda}_2]^{AB}\right],%
\end{align}\label{Gaussal}%
\end{subequations}%
where we have defined the $\mathfrak{sl}(2,\C)$ Lie bracket
\begin{equation}
\ou{[\Lambda_1,\Lambda_2]}{A}{B}=\ou{[\Lambda_1]}{A}{C}\ou{[\Lambda_2]}{C}{B}-\ou{[\Lambda_2]}{A}{C}\ou{[\Lambda_1]}{C}{B}.\end{equation}
The vector constraint, on the other hand, gives rise to two-dimensional diffeomorphisms modulo $SL(2,\C)$ gauge transformations. If $N^a$ and $M^a$ denote vector-valued test functions on $\mathcal{S}$, we find after a straightforward calculation that
\begin{align}
&\left\{H_a[N^a],H_b[M^b]\right\}=-H_a[\mathcal{L}_NM^a]+\nonumber\\
&\qquad+\frac{1}{\gamma}\left[G_{AB}\Big[N^aM^b\uepsilon_{ab}\pi^A\ell^B\Big]+\utilde{G}_{AB}\Big[N^aM^b\uepsilon_{ab}\utilde{\pi}^A\utilde{\ell}^B\Big]+\CC\right],\label{vekal}
\end{align}
where $\mathcal{L}_NM^a=[N,M]^a$  is the Lie derivative.
\vspace{1em}

It is instructive to work out and see that $H_a[N^a]$ and $G_{AB}[\Lambda^{AB}]$ are indeed generators of two-dimensional diffeomorphism and local $SL(2,\C)$ gauge transformations. First of all, we have
\begin{subequations}
\begin{align}
&\left\{H_a[N^a],\ell^A\right\}=N^aD_a\ell^A=\mathcal{L}_N^\uparrow\ell^A\label{Lie1},\\
&\left\{H_a[N^a],\sfpi_{A}\right\}=D_a(N^a\sfpi_{A})=\mathcal{L}_N^\uparrow\sfpi_{A}\label{Lie2},\\
&\left\{H_b[N^b],\ou{A}{AB}{a}\right\}=\frac{1}{\gamma}N^b\uepsilon_{ab}\sfpi^{(A}\ell^{B)}.\label{Lie3}
\end{align}\label{VecLie}%
\end{subequations}
Equally for $\utilde{\ell}^A$, $\utilde{\sfpi}_A$ and $\ou{\utilde{A}}{A}{Ba}$. Equation \eref{Lie1} and \eref{Lie2} define the horizontal lift $\mathcal{L}^\uparrow$ of the Lie derivative into the spin bundle. The third Poisson bracket \eref{Lie2}, on the other hand, returns a diffeomorphism only on-shell\,---\,only if, in fact, the Gauss constraint \eref{Gausscons} is satisfied, in which case
\begin{equation}
\frac{1}{\gamma}N^b\uepsilon_{ab}\sfpi^{(A}\ell^{B)}\approx-\frac{1}{\gamma}N^b\uepsilon_{ab}\oepsilon^{cd}\ou{F}{AB}{cd}=N^b\ou{F}{AB}{ba}=\mathcal{L}_N^\uparrow\ou{A}{AB}{a},
\end{equation}
where the symbol \qq{$\approx$} denotes equality modulo terms constrained to vanish.

For the Gauss constraint, the situation is easier. There we get the fundamental transformations
\begin{subequations}
\begin{align}
&\left\{G_{CD}[\Lambda^{CD}],\ell^A\right\}=-\ou{\Lambda}{A}{B}\ell^B,\\
&\left\{G_{CD}[\Lambda^{CD}],\sfpi_A\right\}=+\ou{\Lambda}{B}{A}\sfpi_B,\\
&\left\{G_{CD}[\Lambda^{CD}],\ou{A}{A}{Ba}\right\}=D_a\ou{\Lambda}{A}{B},
\end{align}\label{GaussLie}%
\end{subequations}
which are the generators of right translations along the fibres of the $SL(2,\C)$ principal bundle and the associated spin bundle (with sections $\ell^A$ and $\sfpi_A$). The situation on the other side of the interface with variables $\utilde{\ell}_A$, $\utilde{\sfpi}_A$ and $\ou{\utilde{A}}{A}{Ba}$ is completely analogous.

We then also have the matching constraint \eref{match1}, which Poisson commutes with the $SL(2,\C)$ connections $\ou{A}{A}{Ba}$ and $\ou{\utilde{A}}{A}{Ba}$, but generates the $U(1)_\C$ transformations
\begin{subequations}
\begin{align}
\left\{M[\varphi],\ell^A\right\}&=+\varphi\ell^A,\\
\left\{M[\varphi],\sfpi_A\right\}&=-\varphi\sfpi_A,
\end{align}\label{matchLie}%
\end{subequations}
and equally for $\utilde{\ell}^A$ and $\utilde{\sfpi}_A$. From \eref{matchLie} alone, many Poisson brackets can be inferred immediately, as for instance
\begin{equation}
\left\{M[\varphi],M_a[\bold{J}^a]\right\}=2M_a[\varphi\boldsymbol{J}^a].\label{matchal0}
\end{equation}

Concerning the algebra of constraints, we only need to consider two further Poisson brackets, namely
\begin{align}
\left\{S[N],M_a[\bold{J}^a]\right\}=\frac{2\I}{\beta+\I}\int_{\mathcal{S}}\left[N\bold{J}^a\left(\ell_AD_a\ell^A+\utilde{\ell}_A\utilde{D}_a\utilde{\ell}^A\right)\right],\label{matchal1}
\end{align}
and
\begin{align}\nonumber
\left\{H_a[N^a],M_b[\bold{J}^a]\right\}=\int_{\mathcal{S}}\Big[N^aD_a\ell_A\bold{J}^bD_b\ell^A-D_a\left(\bold{J}^a\ell_A\right)N^bD_b\ell^A+\\
-N^a\utilde{D}_a\utilde{\ell}_A\bold{J}^b\utilde{D}_b\utilde{\ell}^A+\utilde{D}_a\left(\bold{J}^a\utilde{\ell}_A\right)N^b\utilde{D}_b\utilde{\ell}^A\Big].
\end{align}
If we perform a partial integration ($N^a$ and $\bold{J}^a$ all have compact support) and bring all covariant derivatives on one side, this can be simplified to
\begin{align}\nonumber
&\left\{H_a[N^a],M_b[\bold{J}^a]\right\}=
-M_a\left[\mathcal{L}_N\bold{J}^a\right]+\\
&\quad+\frac{1}{\gamma}\left[G_{AB}\Big[N^a\bold{J}^a\uepsilon_{ab}\ell^A\ell^B\Big]+
\utilde{G}_{AB}\Big[N^a\bold{J}^a\uepsilon_{ab}\utilde{\ell}^A\utilde{\ell}^B\Big]
\right],\label{matchal2}
\end{align}
where the Lie derivative of the vector-valued density $\bold{J}^a$ on $\mathcal{S}$ is defined by
\begin{equation}
\mathcal{L}_N\bold{J}^a=2D_b\big(N^{[b}\bold{J}^{a]}\big)+N^aD_b\bold{J}^b.
\end{equation}

The equations \eref{Gaussal}, \eref{vekal}, \eref{matchal0}, \eref{matchal1} and \eref{matchal2} give already all relevant Poisson brackets. All other Poisson brackets among the constraints can be inferred trivially from either \eref{VecLie} or \eref{GaussLie}. So as for e.g.
\begin{equation}
\left\{H_a[N^a],S[M]\right\}=-S[\mathcal{L}_NM],
\end{equation}
which is an immediate consequence of \eref{VecLie}.

\paragraph{First class and second class constraints} We have now collected all Poisson brackets that are necessary to identify the first class and second class constraints of the system. The vector constraint $H_a[N^a]$ (see \ref{VecLie}) generating two-dimensional diffeomorphisms on $\mathcal{S}$ is first class. And so are the Gauss constraints $G_{AB}[\Lambda^{AB}]$ and $\utilde{G}_{AB}[\utilde{\Lambda}^{AB}]$, which generate $SL(2,\C)$ gauge transformations \eref{GaussLie} on either side of the interface. Equally, for the matching constraint: $M[\varphi]$ is first class and generates the $U(1)_\C$ transformations \eref{matchLie} of the boundary spinors. 

We are then left with the scalar constraint $S[N]$ (as in \ref{Skalarcons}) and the matching condition $M_a[\bold{J}^a]$ (as in \ref{match2}). The scalar constraint is second class, which is a consequence of \eref{matchal2}. For the matching condition, the situation is more complicated. The constraint $M_a[\bold{J}^a]$ is complex-valued, hence there are $2\times 2=4$ such conditions per point. One of them is second class, all others are again first class. This can be seen as follows: In general, we will have that $\ell_AD_a\ell^A$ does not vanish\footnote{If $\ell_AD_a\ell^A=0$, we have one further constraint, which renders the entire system of constraints (\ref{Gausscons}, \ref{Skalarcons}, \ref{Vekcons}, \ref{match1}, \ref{match2}) first-class.} on phase space. We can then parametrise the densitised vector $\bold{J}^a$ as follows
\begin{align}
\bold{J}^a(z^{(\mathnormal{0})},z^{(\mathnormal{1})})&=z^{\mathnormal{(0)}}\oepsilon^{ab}\ell_AD_b\ell^A+
z^{(\mathnormal{1})}d^2v_{\mathnormal{0}}\, {}^{\mathnormal{0}\!}q^{ab}(\beta+\I)\bar{\ell}_{\bar A}D_b\bar{\ell}^{\bar A},
\end{align}
where ${}^{\mathnormal{0}\!}q^{ab}$ is a fiducial signature $(+$$+)$ two-metric on $\mathcal{S}$ and $d^2v_{\mathnormal{0}}$ denotes the corresponding area element. The only relevant Poisson bracket for determining the second class component of $M_a[\bold{J}^a]$ is \eref{matchal2}. For generic $\ell_AD_a\ell^A\neq 0$ it then suffices to determine those values for $z^{(\mathnormal{0})}$ and $z^{(\mathnormal{1})}$ for which the Poisson bracket $\{M_a[\bold{J}^a(z^{(\mathnormal{0})},z^{(\mathnormal{1})})],S[N]\}$ vanishes \emph{on-shell}\footnote{Given all constraints (\ref{Gausscons}, \ref{Skalarcons}, \ref{Vekcons}, \ref{match1}, \ref{match2}) are satisfied.} for all $N$. Going back to \eref{matchal1}, we see that this is certainly true for all $z^{(\mathnormal{1})}=0$. In other words
\begin{equation}
\left\{M_a\Big[
\bold{J}^a(z^{(\mathnormal{0})},0)\Big],S[N]\right\}\approx 0,
\end{equation}
where the symbol \qq{$\approx$} denotes equality modulo terms constrained to vanish. All other relevant Poisson brackets involving $M_a[\bold{J}^a(z^{(\mathnormal{0})},0)]$ weakly vanish as well, which is an immediate consequence of \eref{VecLie}, \eref{GaussLie} and \eref{matchLie}. Hence the constraint
\begin{equation}
M_a\Big[
\bold{J}^a(z^{(\mathnormal{0})},0)\Big]\label{fclass1}
\end{equation}
is first class. When $z^{(\mathnormal{1})}=\bar{z}^{(\mathnormal{1})}=x^{(\mathnormal{1})}$ is real, we get an additional first class constraint, namely
\begin{equation}
M_a\Big[
\bold{J}^a(0,x^{(\mathnormal{1})}))\Big]+\CC\label{fclass2}
\end{equation}
That \eref{fclass2} defines another first class constraint follows from \eref{matchal1} and 
\begin{equation}
\left\{M_a\Big[
\bold{J}^a(0,x^{(\mathnormal{1})}))\Big],S[N]\right\}\approx 0.\label{weaklyzero}
\end{equation}
All other constraints Poisson commute with \eref{fclass2}. This can be inferred already from the infinitesimal gauge transformations \eref{VecLie}, \eref{GaussLie} and \eref{matchLie}.

We are thus left to identify the single second class component of $M_a[\bold{J}^a]$. It is given by the expression
\begin{equation}
M_a\Big[
\bold{J}^a(0,\I y^{(\mathnormal{1})})\Big]+\CC,\label{sclass}
\end{equation}
where $z^{(\mathnormal{1})}=\I y^{(\mathnormal{1})}$ is purely imaginary. That this defines a second class constraint, follows from \eref{matchal1} through
\begin{align}\nonumber
\left\{M_a\Big[
\bold{J}^a(0,\I y^{(\mathnormal{1})}))\Big],S[N]\right\}&+  \CC\approx \\
&\approx4\int_{\mathcal{S}}d^2v_{\mathnormal{0}}\,{}^{\mathnormal{0}\!}q^{ab}\ell_AD_a\ell^A\bar\ell_{\bar A}D_b\bar\ell^{\bar A}\stackrel{\text{i.g.}}{\neq}0,
\end{align}
which does not vanish unless $\ell_AD_a\ell^A=0$.

\paragraph{Gauge symmetries} We have now identified all first class constraints of the system: The two Gauss constraints $G_{AB}[\Lambda^{AB}]$ and $\utilde{G}_{AB}[\utilde{\Lambda}^{AB}]$ on either side of the interface, the vector constraint \eref{Vekcons} and the matching condition \eref{match1}, which generate two copies of internal $SL(2,\C)$ gauge transformations (on either side of the interface), two-dimensional diffeomorphisms of $\mathcal{S}$ and $U(1)_\C$ gauge transformations \eref{matchLie}. We then have additional first-class constraints, which can be identified with those components \eref{fclass1} and \eref{fclass2} of $M_a[\bold{J}^a]$ that Poisson commute with the scalar constraint $S[N]$. The geometric meaning of local $SL(2,\C)$ gauge transformations, $U(1)_\C$ transformations and two-dimensional diffeomorphisms is clear\footnote{The origin of the $U(1)_\C$ gauge symmetry is the invariance of the parametrisation $\varphi^\ast\Sigma_{AB}=\bold{\eta}_{(A}\ell_{B)}$ of the self-dual area two-form $\Sigma_{AB}$ under the $U(1)_\C$ transformations $\ell^A\rightarrow\E^{z/2}\ell^A$ and $\bold{\eta}_A\rightarrow\E^{-z/2}$, see equation \eref{fluxparam} and \eref{pispindef}.}, but what kind of gauge transformations
are generated by those components of $M_a[\bold{J}^a]$ that are first class?

The answer is hidden in an additional gauge symmetry, which appears in the bulk. The bulk action \eref{LBFactn} has in fact more symmetries than just four-dimensional diffeomorphisms and local $SL(2,\C)$ transformations. It enjoys a further gauge symmetry, which renders the entire theory topological. The action is invariant under the infinitesimal shifts 
\begin{subequations}
\begin{align}
\delta_\phi\ou{A}{A}{B}&=-\frac{1}{\gamma}\ou{\phi}{A}{B},\label{shiftsymm1}\\
\delta_\phi\Pi_{AB}&=+\frac{1}{2}D\phi_{AB},
\end{align}\label{shiftsymm}%
\end{subequations}
where the difference tensor $\ou{\phi}{A}{B}$ is an $\mathfrak{sl}(2,\C)$ valued one-form in the bulk. The addition of the boundary action breaks this \emph{shift symmetry}, but only partially. To understand this more explicitly, let us first define
\begin{equation}
\sfPi_{AB}=\varphi^\ast_t\Pi_{AB}\equiv\frac{1}{2}\oepsilon^{ab}\left[\varphi^\ast_t\Pi_{AB}\right]_{ab},
\end{equation}
which is the pull-back of the canonical bulk momentum $\Pi_{AB}$ to a $t=\mathrm{const}.$ slice $\mathcal{S}_t$ in the boundary component $\mathcal{N}$ of $\mathcal{M}$. Now, the boundary spinors $\bold{\pi}_A$ and $\ell_A$ are eigen-spinors of the pull-back of $\Pi_{AB}$ to $\mathcal{N}$(see equation \eref{Psourced} above). We thus also know that 
\begin{equation}
\sfPi_{AB}=\frac{1}{2}\sfpi_{(A}\ell_{B)},
\end{equation}
which is the pull-back of equation \eref{Psourced} down to a $t=\mathrm{const}.$ slice $\mathcal{S}_t$. We now want to illustrate that $M_a[\bold{J}^a]$ generates a version of the shift symmetry \eref{shiftsymm} 
on the boundary. Using the Poisson brackets \eref{Poiss1} and \eref{Poiss2}, a short calculation gives
\begin{subequations}
\begin{align}
\left\{M_b[\bold{J}^b],\ou{A}{A}{Ba}\right\}&=-\frac{1}{\gamma}\bold{J}^b\uepsilon_{ba}\ell^A\ell_B,\\
\nonumber\left\{M_b[\bold{J}^b],\sfPi_{AB}\right\}&=\frac{1}{2}\left\{M_b[\bold{J}^b],\sfpi_{(A}\ell_{B)}\right\}=\\
&=\bold{J}^a(D_a\ell_{(A})\ell_{B)}+\frac{1}{2}\ell_A\ell_BD_a\bold{J}^a.
\end{align}\label{shiftgen}%
\end{subequations}
Notice that the last line is nothing but the covariant exterior derivative of the first. Comparison with \eref{shiftsymm} allows us then to match the gauge parameter $\ou{\phi}{A}{Ba}\in\Omega^1(\mathcal{M}:\mathfrak{sl}(2,\C))$ for the shift symmetry in the bulk, with the gauge parameter $\bold{J}^a$ at the boundary by demanding that
\begin{equation}
\left[\varphi^\ast_t\phi_{AB}\right]_a=\bold{J}^b\uepsilon_{ba}\ell_A\ell_B.\label{gaugeparam}
\end{equation}
The matching constraint $M_a[\bold{J}^a]$ therefore generates a shift transformation \eref{shiftsymm} with gauge parameter \eref{gaugeparam}.	Notice, however, that this shift symmetry is broken partially by the addition of the reality conditions \eref{Skalarcons} at the boundary. Only if $\bold{J}^a$ is of the particular form of \eref{fclass1} or \eref{fclass2}, do we get a symmetry preserving the constraint hypersurface. For generic values of $\bold{J}^a$, the conditions \eref{fclass1} and \eref{fclass2} will be violated, and the Hamiltonian vector field $\{M_a[\bold{J}^a],\cdot\}$ will lie transversal to the constraint hypersurface. 

\paragraph{Dimension of the physical phase space} In summary, the system admits four types of gauge constraints, and two second class constraints. The scalar constraint $S[N]$ and the $y^{(\mathnormal{1})}$-component \eref{sclass} of the matching constraint $M_a[\bold{J}^a]$ are second class. We then have the first class constraints, which are the vector constraint $H_a[N^a]$, generating diffeomorphisms of $\mathcal{S}$, the generators $G_{AB}[\Lambda^{AB}]$ and $\utilde{G}_{AB}[{\utilde{\Lambda}}^{AB}]$ of $SL(2,\C)$ gauge transformations on either side of the interface, the matching constraint $M[\varphi]$, generating $U(1)_{\C}$ transformations and the remaining components \eref{fclass1} and \eref{fclass2} of $M_a[\bold{J}^a]$, which generate the residual shift symmetry \eref{shiftgen} at the boundary. The situation is summarised in the \hyperref[tab 1]{table} below.
\begin{table}[h] 
\begin{tabular}{p{5em}p{17em}p{7.7em}}                      
{Constraints}   &  {Dirac classification} & {DOF removed} \\
\hline      
{$H_a[N^a]$}             & {two first class constraints}  & $2\times 2=4$\\
{$G_{AB}[\Lambda^{AB}]$} & {three $\C$-valued first class constraints} & $2\times 6=12$ \\
{$\utilde{G}_{AB}[\utilde{\Lambda}^{AB}]$} & {three $\C$-valued first class constraints} & $2\times 6=12$ \\
{$M[\varphi]$} & {one $\C$-valued first class constraint} & $2\times 2=4$ \\
{$M_a[\bold{J}^a]$} & {one is second class, three are first class} & $1+2\times 3=7$ \\
{$S[N]$} & {one first class constraint} & $1$ \\
\hline
& & 40\\
\hline
\end{tabular}
\caption{The constraints remove forty dimensions from the phase space of the theory. The phase space has canonical coordinates $\ou{A}{A}{Ba}$, $\sfpi_A$, $\ell^A$ and $\ou{\protect\utilde{A}}{A}{Ba}$, $\protect\utilde{\sfpi}_A$, $\protect\utilde{\ell}^A$, which parametrise  $2\times(6\times 2+2\times 2+2\times 2)=40$ real dimensions. Therefore, all directions in phase space are just pure gauge directions. This renders the boundary theory topological. There are no local degrees of freedom.\label{tab1}}
\end{table}

The counting proceeds as follows: We have two second class constraints and nineteen first class constraints; there are two independent components of the vector constraint, two times six independent constraints generating the $SL(2,\C)$ transformations on either side of the interface, two independent components of the $U(1)_\C$ generators $M[\varphi]$ (the smearing function $\varphi:\mathcal{S}\rightarrow\C$ is complex) and three additional first class constraints, namely \eref{fclass1} and \eref{fclass2} generating the residual shift symmetry \eref{shiftgen} at the boundary. The kinematical phase space, which has canonical\footnote{The symplectic structure is determined by \eref{Poiss1} and \eref{Poiss2}.} coordinates $\ou{A}{A}{Ba}$, $\sfpi_A$, $\ell^A$ and $\ou{\protect\utilde{A}}{A}{Ba}$, $\protect\utilde{\sfpi}_A$, $\protect\utilde{\ell}^A$, has forty dimensions, every first class constraint removes two degrees of freedom, which leaves us with $40-2\times 19=2$ dimensions. There are two additional second class constraints, namely the scalar constraint $S[N]$ and the second class component \eref{sclass} of $M_a[\bold{J}^a]$, which leaves us with no local degrees of freedom along the three-dimensional interface. This  renders the boundary theory topological. All physical degrees of freedom can only appear at the two-dimensional corners.


\subsection{Relevance for quantum gravity}\label{sec4.4} So far, we have only been studying the classical theory. The main motivation concerns, however, non-perturbative approaches to quantum gravity, such as, in particular, loop quantum gravity. Let me explain and justify this expectation, without going into the mathematical details.

Loop quantum gravity can be based either on the phase space \cite{newvariables,ashtekar} for an $SL(2,\C)$ connection or the phase space for an  $SU(2)$ connection \cite{Barberoparam, Immirziparam}. The complex variables have the advantage that local Lorentz invariance is manifest, though we then also need to impose additional reality conditions, which are otherwise already solved implicitly (see e.g.\ \cite{komplex1,Alexandrovanalysis} for a recent analysis on the issue). 

On the phase space for the complex variables, the symplectic structure is determined by the fundamental Poisson brackets for the self-dual variables
\begin{equation}
\left\{\uo{\Pi}{AB}{a}(p),\ou{A}{CD}{b}(q)\right\}=\frac{1}{2}\delta^{(C}_A\delta^{D)}_B\delta^a_b{\delta}^{(3)}(p,q),
\end{equation}
which can be derived from the topological bulk action \eref{LBFactn} as well. 
It was then noted \cite{Bianchi:2009tj, contphas} that the theory can be discretised, or rather truncated, by requiring that the connection be flat everywhere except along the one-dimensional edges $\{\mathcal{E}_i\}$ of a cellular decomposition of the underlying three-manifold\footnote{In a Hamiltonian approach, the three-manifold $\varSigma$ is often required to be a Cauchy hypersurface as in e.g. \cite{twist, komplexspinors, twistintegrals}, a generalisation to null surfaces was proposed as well cf.\ \cite{Mingyitwist}.} $\varSigma$. This results in the cotangent bundle of the moduli space of flat connections on $\varSigma-\bigcup_i\{\mathcal{E}_i\}$, which is a finite-dimensional phase space
\begin{equation}
P_\Gamma=\big(T^\ast SL(2,\C)\big)^L/_\Gamma SL(2,\C)^{N},\label{LQGphasespace}
\end{equation}
built from the cotangent bundle $T^\ast SL(2,\C)$ on each link $l_1,\dots,l_L$ of the graph $\Gamma$ dual to the system of edges $\{\mathcal{E}_i\}$, modulo $SL(2,\C)$ gauge invariance at the nodes $p_1,\dots,p_N$ of the graph. 

The relation to the new boundary variables $\bold{\pi}_A$ and $\ell^A$, which have been introduced in this paper, is as follows. A point in the phase space $T^\ast SL(2,\C)\simeq SL(2,\C)\times\mathfrak{sl}(2,\C)$ of a link is labelled by  a Lie algebra element\footnote{By convention $\Pi^l_{AB}$ is assigned to the fibre over the initial point of the underlying link.} $\Pi_l\in\mathfrak{sl}(2,\C)$, and a link holonomy $h_l\in SL(2,\C)$ connecting the two endpoints. Unless $\Pi^l_{AB}\Pi_l^{AB}=0$, which is incompatible with $\varSigma$ being a spatial hypersurface, the pair $(h_l,\Pi_l)$ can be parametrised by a pair of bi-spinors $(\bar{\pi}^l_{\bar A},\omega^A_l)$ and $(\utilde{\bar\pi}^l_{\bar A},\utilde{\omega}^A_l)$ on either end of the link. The parametrisation is the following
\begin{subalign}
&\Pi_{AB}^l=\frac{1}{2}\pi_{(A}^l\omega_{B)}^l,\label{discreteflux}\\
&\ou{[h_l]}{A}{B}=\frac{\utilde{\omega}^A_l\pi_B^l-\utilde{\pi}^A_l\omega_B^l}{\sqrt{\smash[b]{\pi_C^l\omega^C_l}}\sqrt{\smash[b]{\utilde{\pi}_D^l\utilde{\omega}^D_l}}}\label{discretehol}.
\end{subalign}
One then \emph{postulates} Poisson brackets
\begin{align}
\big\{\pi_A^l,\omega_B^l\big\}=\epsilon_{AB},\quad
\big\{\utilde{\pi}_A^l,\utilde{\omega}_B^l\big\}=-\epsilon_{AB},\label{discretePoiss}
\end{align}
and shows that the symplectic reduction with respect to the \emph{area-matching} constraint
\begin{equation}
M_l=\pi_A^l\omega^A_l-\utilde{\pi}_A^l\utilde{\omega}^A_l,\label{discretematch}
\end{equation}
returns the relevant portion of $T^\ast SL(2,\C)$, which are those parts of $T^\ast SL(2,\C)$ for which the flux is non-degenerate, i.e.\ $\Pi_{AB}^l\Pi^{AB}_l\neq0$. Finally, one needs to impose additional reality conditions, otherwise the self-dual flux $\Pi^l_{AB}$ is incompatible with a four-dimensional metric. One of them is 
\begin{equation}
\frac{\I}{\beta+\I}\pi^l_A\omega^A_l+\CC=0,\label{discreterealty}
\end{equation}
the other requires that $n^{A\bar A}\omega^l_A\bar\pi^l_{\bar A}=0$, where $n^{A\bar A}$ is the spinor equivalent of a normal vector to the face $f_l$ dual to the link. This normal is often required to be timelike, if it is, however, null rather than timelike, the condition simplifies: There is then always a spinor $\ell^A$ on $f_l$, such that $n^{A\bar A}=\I\ell^A\bar\ell^{\bar A}$, which implies that either  $\omega^l_A$ or $\pi_A^l$ must be themselves proportional to $\ell_A$.  The roles of $\pi_A^l$ and $\omega_A^l$ are interchangeable, $\ell^A$ is unique modulo $U(1)_\C$ transformations, and we can, therefore, always restrict ourselves to the case for which $\omega^A=\ell^A$, or more precisely
\begin{equation}
\omega_A^l=\uo{[h_{p\rightarrow l(0)}]}{A}{B}\ell_B(p),\quad \utilde{\omega}^l_A=\uo{[h_{p\rightarrow l(1)}]}{A}{B}\utilde{\ell}_B(p),\label{omident}
\end{equation}
where $p=f_l\cap l$ is the intersection of the link $l$ with its dual face, and $h_{p\rightarrow l(\epsilon)}$ is the parallel transport along the link going from the intersection $p$ towards either endpoint $l(0)$ or $l(1)$. This in turn suggests to identify the conjugate spinors with the two-dimensional surface integrals \begin{equation}
\pi_A^l=\int_{f_l}\uo{[h_{p\rightarrow l(1)}]}{A}{B}\sfpi_B,\quad \utilde{\pi}_A^l=\int_{f_l}\uo{[h_{p\rightarrow l(1)}]}{A}{B}\utilde{\sfpi}_B.
\end{equation}
 over the faces $\{f_l\}$ dual to the links $l_1,\dots, l_L$ of the graph $\Gamma$. Adopting these identifications, implies to view the faces $f_l$ dual to the links as two-dimensional cross sections of three-dimensional null surfaces $\mathcal{N}_{ij}$. For any such face $f_l$ there is then a three-dimensional internal null boundary $\mathcal{N}_{(ij)(l)}\supset f_l$, and the two pairs of  spinors $(\pi_A^{l},\omega^A_{l})$, $(\utilde{\pi}_A^{l},\utilde{\omega}^A_{l})$ at the endpoints of the link dual to the face correspond to the boundary spinors  $(\bold{\pi}_A, \ell^A)$, $(\utilde{\bold{\pi}}_A, \utilde{\ell}^A)$ on either side of a null interface $\mathcal{N}_{(ij)(l)}$ shining out of $f_l$.

In other words, the discrete loop gravity spinors $(\bar\pi_{\bar A}^{l_i},\omega^A_{l_i})$ and  $(\utilde{\bar\pi}_{\bar A}^{l_i},\utilde{\omega}^A_{l_i})$ on a graph with  links $l_1,\dots, l_L$ mirror the continuous boundary spinors 
 $\bold{\pi}^A$, $\ell^A$ on a family of null surfaces shining out of the faces $f_l$ dual to the links of the graph.
The Poisson brackets \eref{discretePoiss}, which were previously {postulated}, can be then derived from the Poisson brackets \eref{Poiss1} on the null surface, and the decomposition \eref{discreteflux}  of the self-dual flux $\Pi_{AB}^l$ in terms of the discrete spinors is analogous to equation \eref{Psourced}. The same happens for the link holonomy; the $SL(2,\C)$ parallel transport \eref{discretehol} along a link is analogous to the $SL(2,\C)$ gauge transformtion \eref{linkhol} across the interface. Equally for the constraints: The area-matching condition and the reality conditions appear both in the discrete theory on a graph (as in \eref{discretematch} and \eref{discreterealty}) and in the three-dimensional boundary theory (as in \eref{match1} and \eref{Skalarcons}).
There is no doubt that this correspondence must be worked out in more detail. So far, I find the analogy encouraging. It suggests that the kinematical structure of loop quantum gravity\,---\,graphs, operators and spin-network functions\,---\,can be all lifted along null surfaces obtaining a fully covariant picture of the dynamics in terms of a topological field theory with defects.   

\section{Summary, outlook and conclusion}
 
\paragraph{Summary} This paper developed a model for discrete gravity in four spacetime dimensions where the only excitations of geometry are carried along curvature defects propagating at the speed of light. The resulting theory has no local degrees of freedom in the bulk, non-trivial curvature is confined to three-dimensional internal boundaries, which represent a system of colliding null surfaces.

The theory is similar to Regge calculus \cite{Reggecalc} and other discrete approaches, such as 't Hooft's model of locally finite gravity \cite{thooftdefect}, causal dynamical triangulations and causal sets \cite{lollreview,Sorkin:2003bx}, 
but there are fundamental differences. First of all, and most crucially, we have a field theory for the Lorentz connection rather than a lattice model for the metric. This field theory is topological and 
the underlying spacetime manifold splits into a union of four-dimensional cells $\{
\mathcal{M}_i\}:\mathcal{M}=\bigcup_{i=1}^N\mathcal{M}_i$, whose geometry is either flat or constantly curved inside (depending on the value of the cosmological constant $\Lambda$). In every such four-cell, there are no local degrees of freedom. Non-trivial curvature is confined to internal boundaries $\mathcal{N}_{ij}=\mathcal{M}_{i}\cap\mathcal{M}_j$, which are three-dimensional. 

The underlying action \eref{fullactn} is local, and splits into a sum over all four-dimensional building blocks, inner three-boundaries and two-dimensional corners. The internal boundary terms are necessary to have a well-posed variational principle. The problem of finding the correct dynamics for the curvature defects boils then down to finding the right boundary action, which cancels the connection variation from the bulk and consistently glues the cells $\{\mathcal{M}_i\}$ across their boundaries. 

What is then the right boundary term? We are viewing gravity as a Yang\,--\,Mills gauge theory for the Lorentz group. At a boundary, a Yang\,--\,Mills gauge connection couples naturally to its boundary charges. Consider, for example, a configuration where the Yang\,--\,Mills electric field is squeezed into a Wilson line. Wherever this Wilson line ends and hits a two-dimensional boundary, a colour charge appears that cancels the gauge symmetry from the bulk. For an $SL(2,\C)$ Lorentz connection, the relevant charge is spin, which suggests to look for an action with spinors as the fundamental boundary variables. We proposed such an action in \hyperref[sec2]{section 2} for a boundary that is null. That the internal boundaries are null rather than space-like or time-like is well desired, it imposes a local notion of causality: The field strength of the $SL(2,\C)$ connection is trivial everywhere except at the internal boundaries, which are null and represent, therefore, the world sheets of curvature defects propagating at the universal speed of light. This was further justified in \hyperref[sec3.4]{section 3.4}, where we gave a constructive proof of existence: We showed that there are explicit solutions of the equations of motion derived from the action \eref{fullactn}, which represent impulsive gravitational waves. These are exact solutions of Einstein's equations in the neighbourhood of an interface, and may describe e.g.\ the gravitational field of a massless point particle \cite{Aichelburg1971}.

The model is specified by the action for the internal boundaries. This action assumes a surprisingly simple form. It defines, in fact, nothing but the symplectic structure for a spinor $\ell^A$ and its canonical momentum, which is (in three dimensions) a spinor-valued two-form $\bold{\pi}_A$. The resulting boundary action is
\begin{equation}
\int_{\mathcal{N}}\bold{\pi}_A\wedge D\ell^A+\CC\label{bndryterm}
\end{equation}
The configuration spinor $\ell^A$ and the momentum spinor $\bold{\pi}_A$ have an immediate geometric interpretation: the bilinear $\I\ell^A\bar{\ell}^{\bar A}$ defines the null generator of the interface, the spin (1,0) component $\bold{\pi}_{(A}\ell_{B)}$ returns the pull-back of the self-dual component of the Pleba\'nski two-form (\ref{selfdualgamma}, \ref{Psourced}, \ref{twsting}), and the spin $(0,0)$ component $\bold{\pi}_A\ell^A$ defines the two-dimensional area element \eref{areaform}. The spinors at the interface are not completely independent, they are subject to certain constraints. First of all, we have the reality conditions \eref{realcond},  that ensure that the self-dual two-form $\bold{\pi}_{(A}\ell_{B)}$ is compatible 
with a signature $(0$$+$$+$) metric at the interface. Furthermore, there are the glueing conditions \eref{matchcond} that match the intrinsic three-geometry {across} the interface. The only metric discontinuity is in the transversal direction. Adding the glueing conditions to the action has a further effect: Given a boundary metric, the spinors $\ell^A$ and $\bold{\pi}_A$ are unique only up to $SL(2,\C)$ transformations and residual $U(1)_\C$ transformations $\ell^A\rightarrow\E^{z/2}\ell^A$ and $\bold{\pi}_A\rightarrow\E^{-z/2}\bold{\pi}_A$. Clearly, the boundary term \eref{bndryterm} is $SL(2,\C)$ invariant, but it violates this additional $U(1)_\C$ gauge symmetry. The symmetry is restored by the area-matching condition \eref{areamatch}, which is added to the action by replacing the covariant $SL(2,\C)$ derivative $D$ by the $SL(2,\C)\times U(1)_\C$ derivative $D-\omega$, where the $U(1)_\C$ gauge connection $\omega$ acts as a Lagrange multiplier for the constraint to be imposed.

A more thorough analysis of the gauge symmetries was performed in \hyperref[sec4]{section 4}. First of all, we noticed that the action in the bulk \eref{LBFactn} is topological. We then integrated out the self-dual two-form $\Pi_{AB}$ in the bulk obtaining the $SL(2,\C)$ Chern\,--\,Simons action \eref{CSactn} at the three-dimensional interfaces. Next, we performed a $2+1$ split of the boundary action and identified the symplectic structure of the theory.
All fundamental variables appear twice, because every such interface $\mathcal{N}_{ij}$ bounds two bulk regions $\mathcal{M}_i$ and $\mathcal{M}_j$, which induce boundary variables from either side. We found that the $SL(2,\C)$ connection becomes Poisson non-commutative at the boundary, while the configuration spinor $\ell^A$ has the canonically conjugate variable $\bold{\sfpi}_A$, which is the pull-back of the momentum spinor $\bold{\pi}_A$ to a $t=\mathrm{const}.$ hypersurface (equally for $\utilde{\ell}^A$ and $\utilde{\sfpi}^A$ as in \eref{Poiss1} and \eref{Poiss2} above). Finally, we found the canonical Hamiltonian \eref{canHam}, which is a sum over the constraints of the system, which consist of the vector constraint generating two-dimensional diffeomorphisms, a pair of $SL(2,\C)$ Gauss constraints generating $SL(2,\C)$ gauge transformations on either side of the interface, the area-matching condition generating $U(1)_\C$ transformations of the boundary spinors, and finally the three first-class components (\ref{fclass1}, \ref{fclass2}) of the glueing conditions \eref{match2}, which generate the residual shift symmetry (\ref{shiftsymm}, \ref{shiftgen}). All of these constraints are first-class, the reality condition \eref{Skalarcons}, on the other hand, is second class, and so is the fourth component \eref{sclass} of the glueing condition $M_a[\bold{J}^a]=0$. The symplectic reduction removes, therefore, forty dimensions from the kinematical phase space, which is forty-dimensional as well. This brought us to the conclusion that the theory has no local degrees of freedom, neither in the bulk nor at the three-dimensional internal boundaries.

\paragraph{Relevance for quantum gravity} The proposal defines a topological gauge theory with defects. Solutions of the equations of motivation represent distributional spacetime geometries, where the gravitational field is trivial in four-dimensional causal cells, whose boundary is null. The geometry is discontinuous across these internal boundaries, which represent curvature defects propagating at the speed of light.

Our main motivation concerns possible applications for non-perturbative approaches to quantum gravity, such as loop quantum gravity. We have a few indications supporting this idea: First of all, the model has a kinematical phase space, whose canonical structure is extremely close to recent developments in loop quantum gravity. In \cite{twist,Dupuis:2012vp,komplexspinors,twistintegrals,Bianchi:2016hmk,Borja:2010rc}, a new representation of loop quantum gravity has been introduced with spinors as the fundamental configuration variables. This construction was bound to the discrete phase space on a graph. 
An interpretation was missing for what these spinors are in the continuum. This paper closes this gap and provides a continuum interpretation: The loop gravity spinors are the canonical boundary degrees of freedom of the gravitational field on a null surface.

The most interesting indication in favour of our proposal concerns its dynamical structure. The theory is topological and this suggests that the transition amplitudes, which are formally given by the path integral
\begin{align}%
\mathcal{Z}_{\{\mathcal{M}_i\}}&[\Psi]=\int\mathcal{D}[\Pi,A,\bold{\pi},\ell]\,\Delta_{\text{FP}}\,\delta(\text{constraints})\,\Psi[A,\ell]\times\nonumber\\\nonumber
\noindent &\times\prod_{\{\mathcal{M}_{i}\}}\exp\bigg(\frac{2\I}{\hbar}\int_{\mathcal{M}_i}\bigg[\Pi_{AB}\, F^{AB}+\frac{1}{\gamma}\Pi_{AB}\,\Pi^{AB}\Big]+\CC\bigg)\times\\	
&\times\prod_{\{\mathcal{N}_{ij}\}}\exp\bigg(\frac{\I}{\hbar}\int_{\mathcal{N}_{ij}}
\bold{\pi}_AD\ell^A+\CC\bigg),\label{pathint}
\end{align}
for fixed boundary states $\Psi[A,\ell]$ in a boundary Hilbert space $\mathcal{H}_{\partial\mathcal{M}}$ exist and  turn into ordinary integrals (or sums) over the moduli of the theory. This would be reminiscent of quantum gravity in three dimensions, where the Ponzano\,--\,Regge amplitudes can be written as a product over $SU(2)$ group integrals for each edge times $SU(2)$ delta functions imposing the flatness of the connection (see \cite{Wieland:2014ab} for a recent derivation). Such moduli exist, and the simplest example is the four-volume\footnote{That the four-volume ${}^4\mathrm{Vol}[\mathcal{M}_i]$ defines an observables is straightforward to see: Clearly, it is invariant under local Lorentz transformations and diffeomorphisms that preserve the bulk region $\mathcal{M}_i$. It is also invariant under the residual shift symmetry \eref{shiftsymm}. A shift symmetry generates a variation $\delta_\phi\Sigma_{AB}\propto D\phi_{AB}$. The equation of motion $D\Sigma_{AB}=0$ (as in \eref{torsless} for $\Pi_{AB}\propto\Sigma_{AB}$), implies that the gauge variation of the four-volume $\delta_\phi{}^4\mathrm{Vol}[\mathcal{M}_i]$ yields a bulk integral over a total exterior derivative, which turns into a boundary integral $\propto\int_{\partial\mathcal{M}_i}\phi_{AB}\wedge\Sigma^{AB}$ that vanishes due to the boundary conditions \eref{gaugeparam} for the gauge parameter $\phi_{AB}$, i.e.\ $\phi_{AB}|_{\partial\mathcal{M}}\propto\ell_A\ell_B$, which implies, in turn, $\phi^{AB}\Sigma_{AB}\propto\ell^A\ell^B\Sigma_{AB}\propto \ell^A\ell^B\bold{\pi}_A\ell_B=0$ at the boundary.}
\begin{equation}
{}^4\mathrm{Vol}[\mathcal{M}_i]=\frac{\I}{3!}\int_{\mathcal{M}_i}\Sigma_{AB}\wedge\Sigma^{AB}+\CC,
\end{equation}
 of a given four-cell $\mathcal{M}_i$, or the trace of the $SL(2,\C)$ holonomy around the perimeter of a corner. The existence of such non-local observables is an important hint that the formal definition of the path integral \eref{pathint} has a mathematical precise meaning and defines a so-called spinfoam model, which is given by certain fundamental amplitudes assigned to the $n$-dimensional ($n=0,\dots,4$) building blocks, glued and traced together according to the adjacency relations of the underlying cellular decomposition of the four-manifold $\mathcal{M}$ (such as in three dimensions where the $6j$-symbol defines the vertex amplitude for the Ponzano\,--\,Regge model). 
\paragraph{Perspectives}
The amplitudes \eref{pathint} are defined for a given and fixed family of four-dimensional cells $\{\mathcal{M}_i\}$, which are glued among bounding interfaces $\{\mathcal{N}_{ij}\}$. This combinatorical structure is an ad-hoc input, which enters the classical action \eref{fullactn} as an external background structure. How are then different discretizations $\{\mathcal{M}_i\}$ with different combinatorial structures supposed to be taken into account? There are two possible answers to this question: In the first scenario, the full theory will be defined through a continuum limit, which sends the number of  four-dimensional cells to infinity. The definition of the theory would then most likely include some sort of renormalisation group flow, which would give a prescription for how to take this limit in a rigorous manner. 
The main conceptual difficulty with such an approach is that there is no fundamental lattice scale entering the action \eref{fullactn}. Indeed, it is the gravitational field itself that determines the size of the individual building blocks, and this makes it difficult to identify the correct variables and the correct notion of scale to study the renormalization group flow. Therefore, more sophisticated tools and techniques such as those developed for Regge calculus \cite{Feinberg:1984he,Barrett:1988wd, barrett1988convergence} and so-called spinfoam models \cite{Bahr:2009mc,Bahr:2009qc,Dittrich:2014ala,Dittrich:2011zh} may be required. 

The second possibility, which I find more appealing, is a more radical idea. In this scenario, the amplitudes for a given and fixed configuration of four-cells would be seen as Feynman amplitudes for an auxiliary quantum field theory. To define the entire theory, one would then {sum} over an infinite, but most likely very preferred class of combinatorical structures, which would arise from the perturbative expansion of the auxiliary field theory. The approach would be conceptually very similar to group field theory \cite{Oriti:2014aa,Oriti:2013aa,oritigft}, where the gravitational path integral on a given simplicial discretisation arises from the perturbative expansion of a quantum field theory over a group manifold.
\vspace{0.5em}

Finally, there is one obvious open question that I have avoided altogether, namely how the two physical degrees of freedom of general relativity should come out of the model. This question is certainly related to the previous question regarding the continuum limit, but some hints of an answer should already appear at the level of the microscopic theory, which is defined by the action \eref{fullactn}. This action  was constructed such that the solutions of the equations of motion represent four-dimensional distributional geometries, where the curvature is trivial in four-dimensional cells, which are glued among bounding null-surfaces. The geometry is described in terms of $SL(2,\C)$ gauge variables (an $SL(2,\C)$ connection in the bulk coupled to spinors at the internal null boundaries). If we then take the quotient by the internal $SL(2,\C)$ gauge transformations, we are left with a theory that can only be described by a metric and a connection. The connection satisfies the torsionless condition \eref{torsless}, hence we expect that the only relevant degrees of freedom are captured by the metric, which is now locally flat. We then saw in section \ref{sec3.2} that special solutions of the equations of motion exist that have a non-trivial distributional curvature tensor at the defect: The resulting Weyl tensor is of Petrov type IV, thus describing transverse gravitational radiation. But we then also saw that there are solutions where both the traceless part of the Ricci tensor \eref{Riccispin} and the Weyl tensor \eref{Weylspin}  are non-vanishing, the Ricci tensor being $R_{ab}\sim\ell_a\ell_b$. This could now either mean that the model already includes some sort of distributional matter (as in e.g.\ string theory), with a distributional energy momentum tensor $T_{ab}\sim\ell_a\ell_b$, or\,---\,and I find this more likely\,---\,that the model describes a metric theory of gravity with more than just two propagating degrees of freedom. A minimal example for such a theory is given by the Starobinsky model \cite{Kehagias:2014aa} of inflation, which has three propagating degrees of freedom (which are given by the two polarisations of gravitational radiation, and one additional spin-$0$ scalar mode). I find this idea very promising and exciting, and it is, in fact, the line of reasoning that I am currently investigating. A more rigorous analysis will be presented in an upcoming article, which is currently under preparation. 

\paragraph{Acknowledgments}

This research was supported in part by Perimeter Institute for Theoretical Physics. Research at Perimeter Institute is supported by the Government of Canada through the Department of Innovation, Science and Economic Development and by the Province of Ontario through the Ministry of Research and Innovation.

\appendix
\renewcommand*{\thesection}{\Alph{section}}
\section*{Appendix: Spinors and world tensors}\label{spinappdx}\setcounter{section}{1}
Following Penrose's notation, we write $\ell^A$ with $A,B,C,\dots$ to denote a two-component spinor that transforms under the fundamental representation of $SL(2,\C)$, primed indices $\b{A},\b{B},\b{C},\dots$ refer to the complex conjugate representation. The indices are raised and lowered using the anti-symmetric epsilon tensor $\epsilon_{AB}=-\epsilon_{BA}$, which commutes with the group action. Our conventions are
\begin{equation}
\ell_A=\epsilon_{BA}\ell^B,\quad\ell^{A}=\epsilon^{AB}\ell_B,\quad
\b\ell_{\b A}=\b\epsilon_{\b B\b A}\b\ell^{\b B},\quad\b\ell^{\b A}=\b\epsilon^{\b A\b B}\b\ell_{\b B},
\end{equation}
with $\epsilon^{AC}\epsilon_{BC}=\delta^A_B$. The relation between spinors and internal Minkowski vectors $v^\alpha$ is provided by the soldering form $\ou{\sigma}{A\b A}{\alpha}$. The generalised Pauli matrices provide an explicit matrix representation
\begin{equation}
\begin{pmatrix}
\ou{\sigma}{0\b 0}{0}&\ou{\sigma}{0\b 1}{0}\\
\ou{\sigma}{1\b 0}{0}&\ou{\sigma}{1\b 1}{0}\\
\end{pmatrix}=
\mathds{1},\quad
\begin{pmatrix}
\ou{\sigma}{0\b 0}{i}&\ou{\sigma}{0\b 1}{i}\\
\ou{\sigma}{1\b 0}{i}&\ou{\sigma}{1\b 1}{i}\\
\end{pmatrix}=
\sigma_i,
\label{soldform}
\end{equation}
where $\mathds{1}$ is the identity matrix and $(\sigma_1,\sigma_2,\sigma_3)$ are the three-dimensional spin matrices. The soldering form $\ou{\sigma}{A\b A}{\alpha}$ maps an internal Lorentz vector $v^\alpha\in\R^4$ into an anti-hermitian\footnote{That $v^{A\b A}$ is anti-hermitian is a consequence of our choice ($-$$+$$+$$+$) for the metric signature.} $2\times 2$ matrix $v^{A\b A}$: $v^\alpha\mapsto v^{\b AA}={\I}/{\sqrt{2}}\ou{\sigma}{A\b A}{\alpha}v^\alpha$. We have $v^\alpha={\I}/{\sqrt{2}}\uo{\b\sigma}{\b A A}{\alpha}v^{A\b A}$ for the inverse map. This isomorphism can be generalised to any world tensor. It maps the Lorentz invariant Minkowski metric $\eta_{\alpha\beta}$ into the product of the $SL(2,\C)$ invariant epsilon tensors $\eta_{\alpha\beta}=\eta_{A\b A B\b B}=\epsilon_{AB}\b\epsilon_{\b A\b B}$ and any $SL(2,\C)$ transformation $g$ (an element of the double cover of the restricted Lorentz group) into a proper orthochronous Lorentz transformation $\ou{\Lambda}{\alpha}{\beta}\equiv\ou{\Lambda}{A\b A}{B\b B}=\ou{g}{A}{B}\ou{\b g}{\b A}{\b B}\in L_+^\uparrow$.

The soldering forms $\ou{\sigma}{A\b A}{\alpha}$ satisfy important algebraic identities. First of all, we have the generalised Pauli identity
\begin{equation}
\ou{\sigma}{A\b C}{\alpha}\b\sigma_{\b CB\beta}=-\delta^A_B\eta_{\alpha\beta}-2\ou{\Sigma}{A}{B\alpha\beta},\label{soldef}
\end{equation}
where
\begin{equation}
\ou{\Sigma}{AB}{\alpha\beta}=\frac{1}{2}\ou{\sigma}{A}{\b C[\alpha}\ou{\b\sigma}{\b CB}{\beta]}
\end{equation}
are the self-dual generators of $SL(2,\C)$. Equation \eref{soldef} implies that the matrices
\begin{equation}
\gamma_\alpha\equiv
\begin{pmatrix}
\ou{[\gamma]}{A}{B\alpha}&\ou{[\gamma]}{A\bar B}{\alpha}\\
{[\gamma]}_{\bar A B\alpha}&\uo{[\gamma]}{\bar A}{\bar B}{}_{\alpha}
\end{pmatrix}=
\begin{pmatrix}
\emptyset&\ou{\sigma}{A\bar B}{\alpha}\\
{\bar\sigma}_{\bar A B\alpha}&\emptyset
\end{pmatrix}
\end{equation}
define a representation of the Clifford algebra $\{\gamma^\alpha,\gamma^\beta\}=-2\eta^{\alpha\beta}\mathds{1}$.
It is then straightforward to check that the self-dual generators $\ou{\Sigma}{A}{B\alpha\beta}$ provide a representation of the commutation relations of the Lorentz group. Indeed, we have
\begin{equation}
\ou{\Sigma}{A}{C\alpha\beta}\ou{\Sigma}{C}{B\mu\nu}-\ou{\Sigma}{A}{C\mu\nu}\ou{\Sigma}{C}{B\alpha\beta}=4\delta^{\alpha^\prime}_{[\alpha}\delta^{\beta^\prime}_{\beta]}\eta_{\beta^\prime\mu^\prime}
\delta^{\mu^\prime}_{[\mu}\delta^{\nu^\prime}_{\nu]}\ou{\Sigma}{A}{B\alpha^\prime\nu^\prime},
\end{equation}
and the matrices $\ou{\Sigma}{A}{B\alpha\beta}$ are self-dual, because
\begin{equation}
\ast\!\Sigma_{AB\alpha\beta}=\frac{1}{2}\uo{\epsilon}{\alpha\beta}{\mu\nu}\Sigma_{AB\mu\nu}=\I\Sigma_{AB\alpha\beta},
\end{equation}
where $\epsilon_{\alpha\beta\mu\nu}$ is the internal Levi-Civita tensor, with conventions $\epsilon_{0123}=1$.

\providecommand{\href}[2]{#2}\begingroup\raggedright\endgroup

\end{document}